# State preservation by repetitive error detection in a superconducting quantum circuit

J. Kelly,[1,*] R. Barends,[1,2,*] A. G. Fowler,[1,3,2,*] A. Megrant,[1,4] E. Jeffrey,[1,2] T. C. White,[1] D. Sank,[1,2] J. Y. Mutus,[1,2] B. Campbell,[1] Yu Chen,[1,2] Z. Chen,[1] B. Chiaro,[1] A. Dunsworth,[1] I.-C. Hoi,[1] C. Neill,[1] P. J. J. O'Malley,[1] C. Quintana,[1] P. Roushan,[1,2] A. Vainsencher,[1] J. Wenner,[1] A. N. Cleland,[1] and John M. Martinis[1,2]

[1]*Department of Physics, University of California, Santa Barbara, CA 93106, USA*
[2]*Present address: Google Inc.*
[3]*Centre for Quantum Computation and Communication Technology, School of Physics, The University of Melbourne, Victoria 3010, Australia*
[4]*Department of Materials, University of California, Santa Barbara, CA 93106, USA*

**Quantum computing becomes viable when a quantum state can be preserved from environmentally-induced error. If quantum bits (qubits) are sufficiently reliable, errors are sparse and quantum error correction (QEC)[1–6] is capable of identifying and correcting them. Adding more qubits improves the preservation by guaranteeing increasingly larger clusters of errors will not cause logical failure – a key requirement for large-scale systems. Using QEC to extend the qubit lifetime remains one of the outstanding experimental challenges in quantum computing. Here, we report the protection of classical states from environmental bit-flip errors and demonstrate the suppression of these errors with increasing system size. We use a linear array of nine qubits, which is a natural precursor of the two-dimensional surface code QEC scheme[7], and track errors as they occur by repeatedly performing projective quantum non-demolition (QND) parity measurements. Relative to a single physical qubit, we reduce the failure rate in retrieving an input state by a factor of 2.7 for five qubits and a factor of 8.5 for nine qubits after eight cycles. Additionally, we tomographically verify preservation of the non-classical Greenberger-Horne-Zeilinger (GHZ) state. The successful suppression of environmentally-induced errors strongly motivates further research into the many exciting challenges associated with building a large-scale superconducting quantum computer.**

The ability to withstand multiple errors during computation is a critical aspect of error correction. We define $n$-th order fault-tolerance to mean that *any* combination of $n$ errors is tolerable. Previous experiments based on nuclear magnetic resonance[8,9], ion traps[10], and superconducting circuits[11–13] have demonstrated multi-qubit states that are first-order tolerant to one type of error. Recently, experiments with ion traps and superconducting circuits have shown the simultaneous detection of multiple types of errors[14,15]. The above hallmark experiments demonstrate error correction in a single round; however, quantum information must be preserved throughout computation using multiple error correction cycles. The basics of repeating cycles have been shown in ion traps[16] and superconducting circuits[17]. Until now, it has been an open challenge to combine these elements to make the information stored in a quantum system robust against errors which intrinsically arise from the environment.

The key to detecting errors in quantum information is to perform QND parity measurements. In the surface code, this is done by arranging qubits in a chequerboard pattern – with data qubits corresponding to the white, and measure qubits to the black squares (see Fig. 1) – and using these ancilla measure qubits to repetitively perform parity measurements to detect bit-flip ($\hat{X}$) and phase-flip ($\hat{Z}$) errors[7]. A square chequerboard with $(4n+1)^2$ qubits is $n$-th order fault tolerant, meaning at least $n+1$ errors must occur to cause failure in preserving a state if fidelities are above a threshold. With error suppression factor $\Lambda > 1$ and more qubits, failure becomes increasingly unlikely with probability $\epsilon \sim 1/\Lambda^{n+1}$ (assuming independent errors). The surface code is highly appealing for superconducting quantum circuits as it requires only nearest neighbour interactions, single and two-qubit gates, and fast repetitive measurements with fidelities above a lenient threshold of approximately 99%. All of this has recently been demonstrated in separate experiments[18,19].

The simplest system demonstrating the basic elements of the surface code is a one-dimensional chain of qubits, as seen in Fig. 1a. It can run the repetition code, a primitive of the surface code, which corrects bit-flip errors on both data and measure qubits. The device shown in Fig. 1b is a chain of nine qubits, which allows us to experimentally test both first- and second-order fault-tolerance. It consists of a superconducting aluminium film on a sapphire substrate, patterned into Xmon transmon qubits[20] with individual control and readout. The qubits are the cross-shaped devices; they are capacitively coupled to their nearest neighbours, controlled with microwave drive and frequency detuning pulses, and measured with a dispersive readout scheme. The device consists of five data qubits and four measure qubits in an alternating pattern, see Supplementary Information for details.

To detect bit-flips, we determine the parity of adjacent data qubits by measuring the operator $\hat{Z}\hat{Z}$. We do this using an ancilla measure qubit, and performing single- and two-qubit quantum gates (Fig. 1c). The operator measurement has the value -1 and leaves the ancilla qubit state unperturbed for states $|00\rangle$ and $|11\rangle$, and value +1 which flips the ancilla qubit state for $|01\rangle$ and $|10\rangle$. Therefore, errors can be detected as they occur by repeating this operator and noting changes in the outcome. Importantly, this measurement does not destroy the quantum nature: given input $\alpha|00\rangle + \beta|11\rangle$ the result is -1 and the quantum state remains, with similar behavior for other Bell-like superposition states. In the repetition code, simultaneous measurements of these operators enable multiple bit-flip errors to be detected.

We now discuss how bit-flip errors, which can occur on any qubit and at any time, are identified. The quantum circuit of the repetition code is shown in Fig. 2a, for three cycles (in time) and nine qubits. This is the natural extension of the schematic in Fig. 1c, optimized for our hardware (Supplementary Information). The figure illustrates four distinct types of bit-flip errors (stars): measurement error (gold), single-cycle data error (purple), two-cycle data error (red), and a data error after the final cycle (blue). Controlled-NOT (CNOT) gates propagate bit-flip errors on the data qubit to the measure qubit. Each of these errors is typically detected at two locations if in the interior and one location if at the boundary; we call these "detection events". The error connectivity graph[21] is shown in Fig. 2b, where the grey lines indicate every possible pattern of detection events that can arise from a single error. The last column of values for the $\hat{Z}\hat{Z}$ operators in Fig. 2b are constructed from the data qubit measurements, so that errors between the last cycle and data qubit measurement can be detected (Supplementary Information).

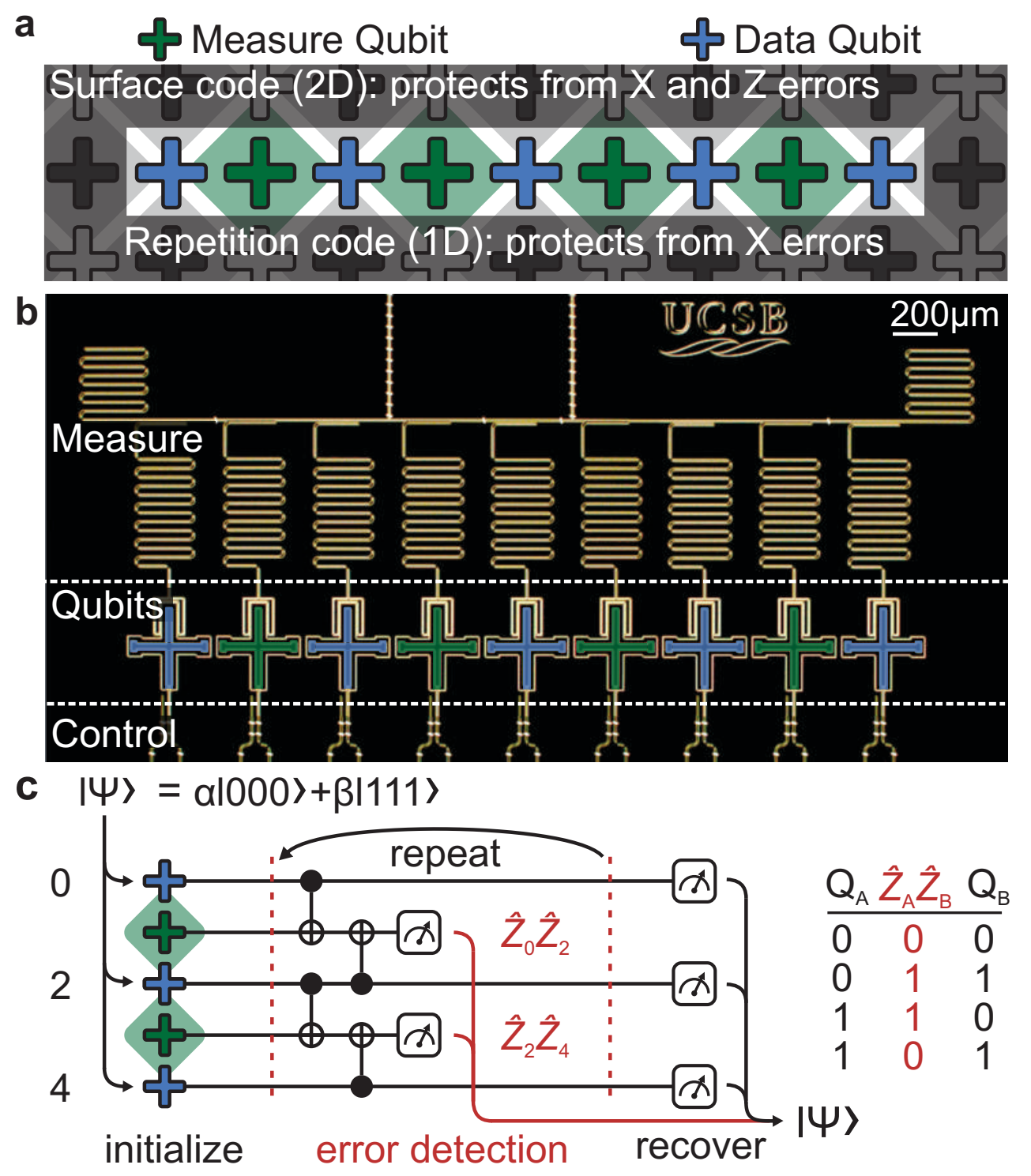

FIG. 1: **Repetition code: device and algorithm.** (a) The repetition code is a one-dimensional variant of the surface code, and is able to protect against $\hat{X}$ (bit-flip) errors. The code is implemented using an alternating pattern of data and measure qubits. (b) Optical micrograph of the superconducting quantum device, consisting of nine Xmon[20] transmon qubits with individual control and readout, with a nearest-neighbour coupling scheme. (c) The repetition code algorithm uses repeated entangling and measurement operations which detect bit-flips, using the parity scheme on the right. Using the output from the measure qubits during the repetition code, the initial state can be protected by detecting physical errors. Measure qubits are initialized into the $|0\rangle$ state and need no reinitialization as measurement is QND.

In the absence of errors, there are two possible patterns of sequential measurement results. If a measure qubit's neighbouring data qubits are in the $|00\rangle$ or $|11\rangle$ state, the measure qubit will report a string of identical values. If the data qubits are in the $|01\rangle$ or $|10\rangle$ state, the measure qubit will report alternating values, as measurement is QND. Single data bit-flip errors make the measurement outcomes switch between these two patterns. For example, if the measurement outcomes for three cycles are 0, 0, and 1, this indicates a change from the identical to the alternating pattern in the last measurement, and hence a detection event. Explicitly, with $m_t$ the measure qubit outcome at cycle $t$ and $\oplus$ the exclusive OR (XOR) operator, for each of the two patterns we have $b_t = m_{t-1} \oplus m_t = 0$ or 1. A detection event at cycle $t$ is then identified when $D_t = b_{t-1} \oplus b_t = 1$.

We use minimum-weight perfect matching[22–24] to decode to physical errors, based on the pattern of detection events and an error model for the system. Intuitively, it connects detection events in pairs or to the boundary using the shortest weighted path length. It is important to note that errors can lead to detection event pairs that span multiple cycles, necessitating the need for multi-round analysis as opposed to round-by-round. See Supplementary Information for details.

To study the ability of our device to preserve quantum states, we initialised the data qubits into a GHZ state $[(|000\rangle +$

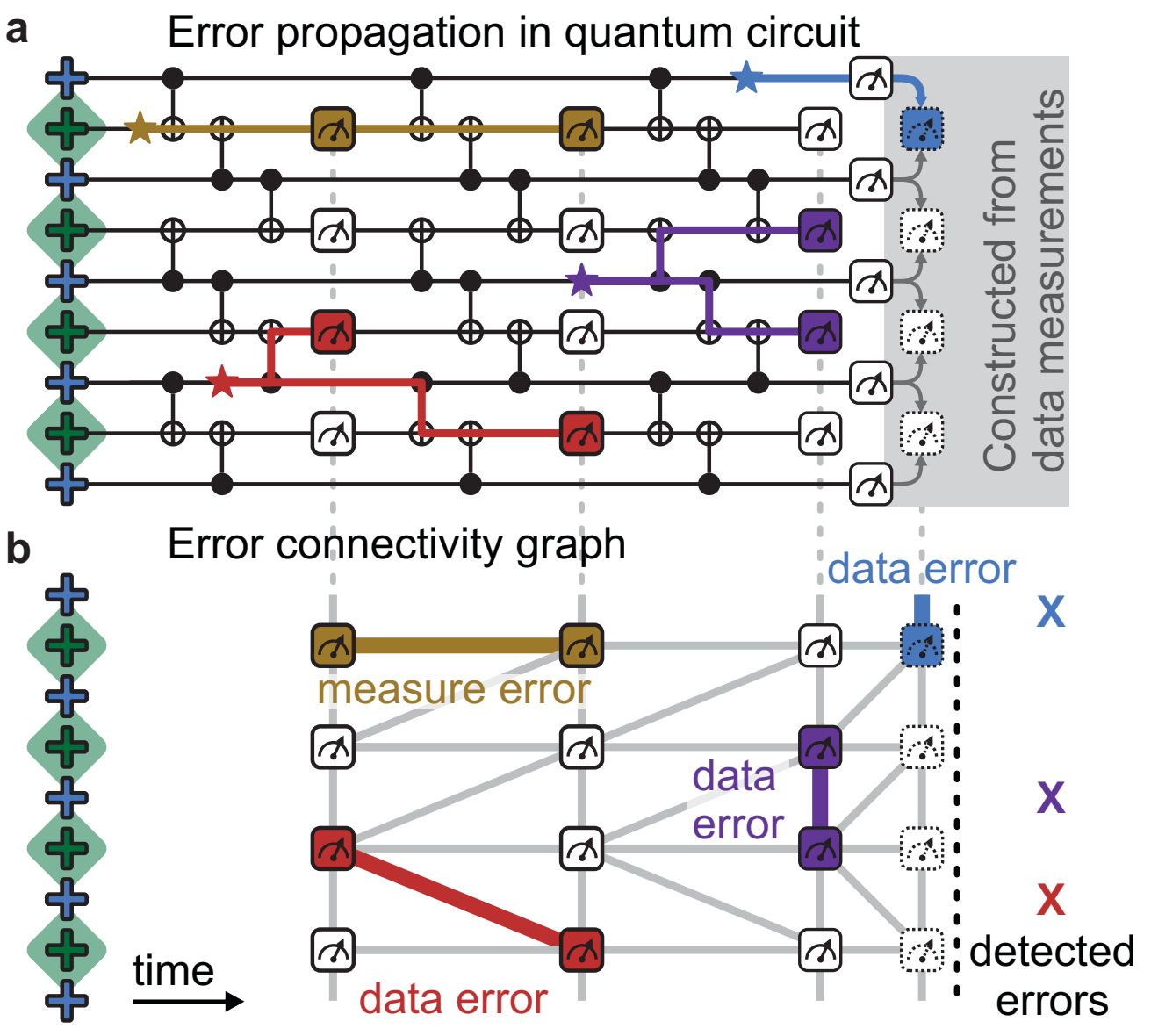

FIG. 2: **Error propagation and identification.** (a) The quantum circuit for three cycles of the repetition code, and examples of errors. Errors propagate horizontally in time, and vertically through entangling gates. Different errors lead to different detection patterns: An error on a measure qubit (gold) is detected in two subsequent rounds. Data qubit errors (purple, red, blue) are detected on neighbouring measurement qubits in the same or next cycle. Data errors after the last round (blue) are detected by constructing the final set of $\hat{Z}\hat{Z}$ eigenvalues from the data qubit measurements. (b) The connectivity graph for the quantum circuit above, showing measurements and possible patterns of detection events (grey), see text. The example detection events and their connections are highlighted, the corresponding detected errors are shown on the right, which when applied, will recover the input data qubit state.

$|111\rangle)/\sqrt{2}]$ and applied two rounds of the repetition code, see Fig. 3. The algorithm is shown in Fig. 3a. Using quantum state tomography we measured the input GHZ state to have a fidelity $\mathrm{Tr}\left(\rho_{\mathrm{ideal}}\rho\right)$ of 82%, above the threshold of 50% for genuine entanglement[25]. After two repetition code cycles, we use tomography to construct the density matrices for each pattern of detection events. We find a state fidelity of 78% in the case of no detection events, indicating a retention of genuine quantum entanglement. In the case of two detection events, which indicate a likely data qubit error in the first cycle, we find elements away from the ideal positions. By applying the recovery operation in postprocessing (a single bit-flip on the blue data qubit) we can restore the state. We find that the off-diagonal elements have not vanished – and genuine entanglement is preserved with a fidelity of 59% – even though the repetition code does not provide phase protection. Reduced but non-zero off-diagonal terms indicate bit errors arise from incoherent processes, like qubit energy relaxation which scrambles the phase, as well as coherent processes. Conditional tomography for every configuration can be found in the Supplementary Information.

The data in Fig. 3 clearly show that the one-dimensional repetition code algorithm does not necessarily destroy the quantum nature of the state. It allows for preserving the quantum state in the case of no errors, and correcting bit-flip errors otherwise. This preservation is achieved purely through error detection and classical post-processing, like for the full surface code, avoiding the need for dynamic feedback with quantum gates.

We now address the critical question of how well our implementation of the repetition code protects logical states over many cycles. The process flow is illustrated in Fig. 4a. We start by initialising the data qubits in either of the logical basis states: $|0_L\rangle = |0..0\rangle$ or $|1_L\rangle = |1..1\rangle$. We then run the repetition code algorithm for $k$ cycles, and finish by measuring the state of all data qubits. We repeat this 90,000 times to gather statistics. The classical measurement results are converted into detection events, which are processed using minimum-weight perfect matching to generate corrections, see Supplementary Information. These corrections are then applied to the measured data qubit output to see if the input is recovered. Due to the topological nature of errors, we either recover the logical state, or the bit-wise inverse (see Supplementary Information). The fidelity of the repetition code algorithm is defined by the success rate of this recovery. In our system, qubits naturally relax to $|0\rangle$, intrinsically making $|0_L\rangle$ more robust than $|1_L\rangle$. To balance these errors and to increase the worst-case lifetime of the system, we apply physical bit-flips to each data qubit at the end of each cycle. This logical flip is compensated for in software. In order to quantify the reduction of logical errors with system size $n$, we have implemented the repetition code with five and nine qubits in total, corresponding to first- and second-order fault-tolerance.

In Fig. 4b we show the fidelity of the repetition code as a function of the number of cycles for five (blue) and nine (red) qubits. We also plot the probability of a $|1\rangle$ state idling for the same duration, averaged over the five data qubits (black). This allows for a direct comparison of single physical qubit error with the multi-qubit logical error. We find a reduced error of logical states after eight cycles as compared to a physical qubit; by a factor of 2.7 for five qubits and 8.5 for nine qubits. We also see a non-exponential fidelity decay for logical states, due to an increasing error rate with cycle number, see Figure 4c. We attribute this to energy relaxation of measure qubits. Initial logical states of all 0's or 1's have even parity for all $\hat{Z}\hat{Z}$ operators, maintaining the initial measure qubit $|0\rangle$ state. A bit-flip error on a data qubit, statistically more likely with increasing cycle number, will cause the nearby $\hat{Z}\hat{Z}$ operators to have odd parity. This will flip measure qubits between the $|0\rangle$ and $|1\rangle$ state at each cycle, making them susceptible to energy relaxation and hence increasing the rate of detection events, see Supplementary Information.

Figure 4 demonstrates state preservation through error correction. We emphasize that we correct errors that intrinsically arise from the environment. Additionally, we see larger repetition codes leading to greater error suppression. This is evidence for the system operating with fidelities above the repetition code threshold. As the error rates grow with cy-cle number, the many-cycle behaviour of the repetition code must be explored to ensure that the the system remains above threshold. The ratio of the errors for the $n = 1$ and $n = 2$ case after eight cycles suggests $\Lambda = 3.2$, but larger system sizes are needed to infer this accurately for large $n$ and verify that the logical error rate is suppressed exponentially as $\epsilon_{\text{logical}} \sim 1/\Lambda^{n+1}$, as desired.

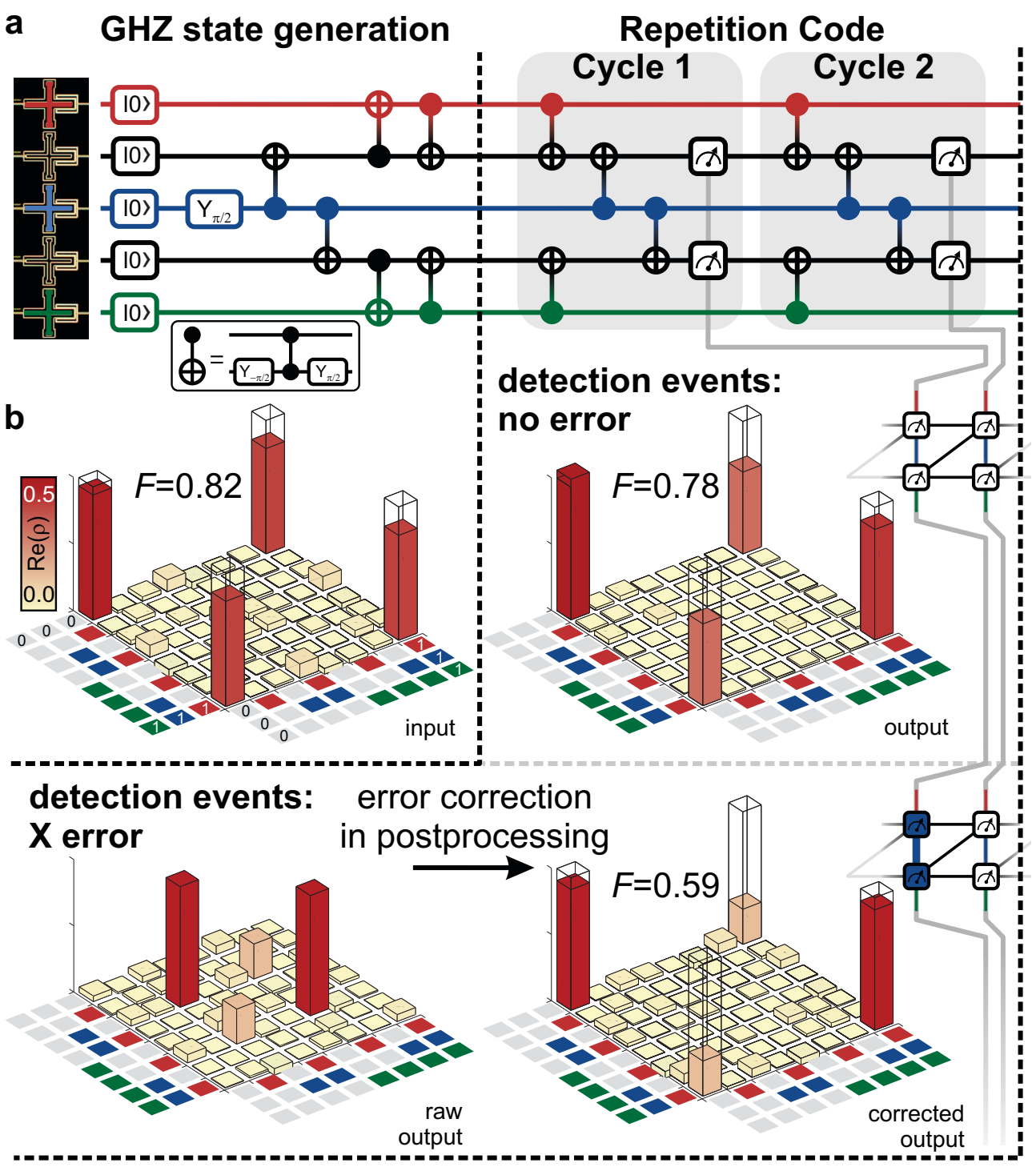

FIG. 3: **Protecting the GHZ state from bit-flip errors.** (a) Quantum circuit for generating the GHZ state and two cycles of the repetition code. CNOT gates are physically implemented with controlled-phase (CZ) and single qubit gates. (b) Quantum state tomography on the input, and after the repetition code conditional on the detection events: We input a GHZ state with a fidelity of 82%, and find, for the case of no detection events, a 78% fidelity GHZ state. For the detection event connecting both measure qubits, indicating a likely bit-flip error on the central data qubit, we find that through correcting in postprocessing by exchanging matrix elements we recover the major elements of the diagonal. We also recover nonzero off-diagonal elements, indicating some bit-flip errors are coherent.

Our demonstration that information can be stored with lower error in logical states than in single physical qubits shows that the basic physical processes required to implement surface code error correction are technologically feasible. We hope our work helps accelerate research into the many outstanding challenges that remain, such as development of two-dimensional qubit arrays, improving gate and measurement fidelities[26], and investigating the many-cycle behavior of error correction schemes.

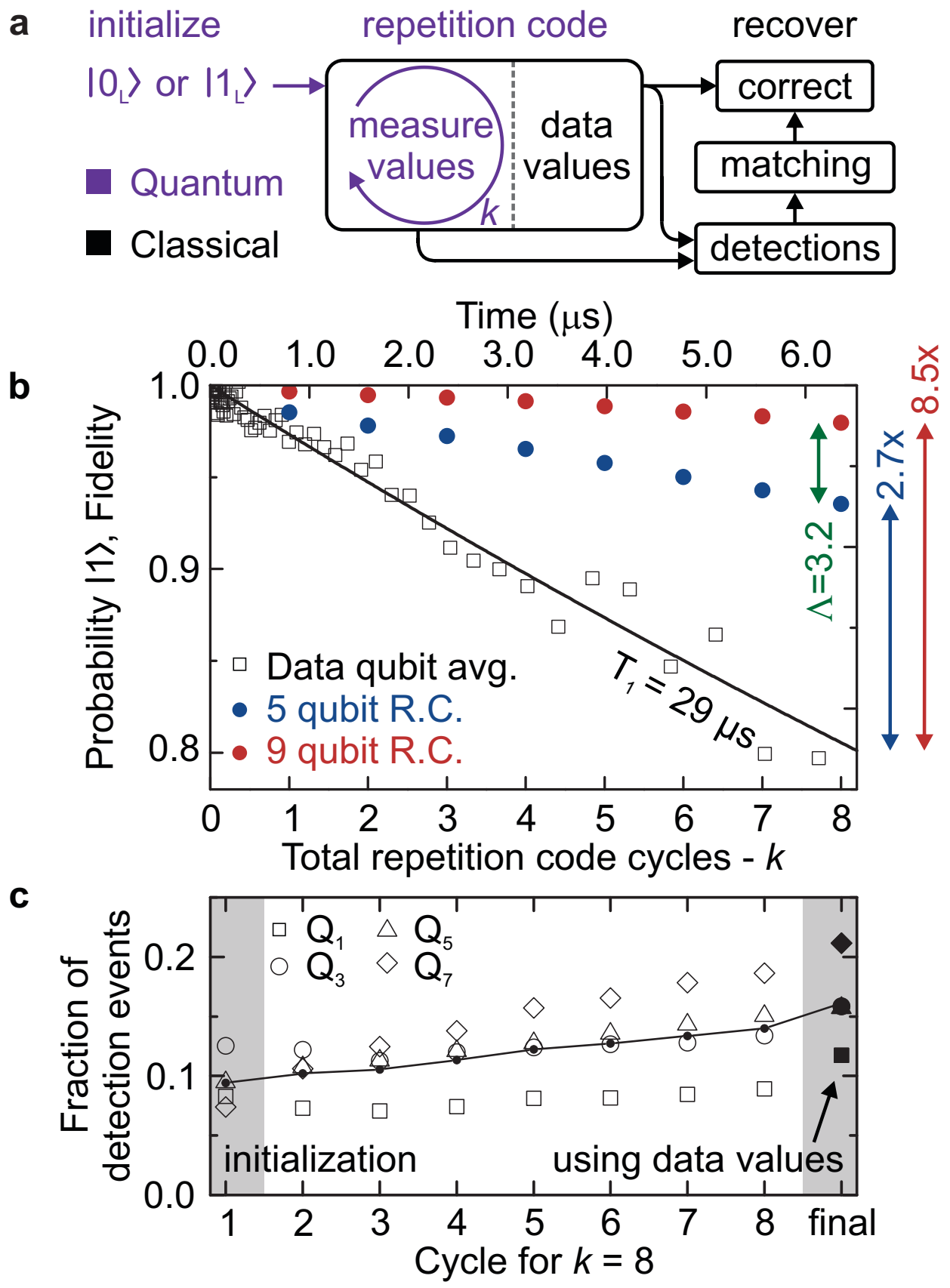

FIG. 4: **Logical state preservation with the repetition code.** (a) Information flowchart of the repetition code. The data qubits are initialised into $|0_L\rangle$ or $|1_L\rangle$, and the repetition code is repeated $k$ times. In post-processing, the measurement qubit outcomes are converted into detection events and matched to find likely errors, see Fig. 2. A successful recovery converts the measured data qubit state into the input state. (b) Memory fidelity $vs.$ time and cycles for a single physical qubit (black) and the five- (blue) and nine- (red) qubit repetition code. Note that energy relaxation decays from a fidelity of 1 to 0, whereas the repetition code decays from a fidelity of 1 to 0.5. Five qubit data sampled from nine qubit data, see Supplementary Information. The average physical qubit lifetime is $T_1 = 29\ \mu$s, and after eight cycles we see an improvement in error rate by a factor of 2.7 for five qubits, and 8.5 for nine qubits when using the repetition code. This indicates a $\Lambda$ parameter of 3.2 for our system after eight cycles. (c) Average number of detection events per measure qubit, $vs.$ cycle number, for experiments consisting of eight cycles. We see an increasing rate of detection events with increasing cycle number. This can be attributed to the statistically increasing number of odd parity $\hat{Z}\hat{Z}$ measurements, see text.

**Methods Summary** The system is brought up in a three step process: characterization, coarse calibration, and fine calibration. 1) Qubit spectra are characterized by analysing raw coherence times at various operating frequencies (Supplementary Information). Using this information, optimal idle, gate, and readout frequencies are chosen. 2) Gates are performed identically to Ref.[18] and optimized as in Ref.[27]. Readout is optimized for maximal separation error while minimizing state transitions. Multi-qubit readout is optimized separately for parametric amplifier saturation[28]. 3) Fine calibration is performed by using the repetition code, where parameters are optimized to reduce detection events.

**Acknowledgements** We thank A. N. Korotkov and D. L. Moehring for useful discussions. We thank P. Duda for help with photomasks and photolithography. This work was supported by the Office of the Director of National Intelligence (ODNI), Intelligence Advanced Research Projects Activity (IARPA), through the Army Research Office grants W911NF-09-1-0375 and W911NF-10-1-0334. All statements of fact, opinion or conclusions contained herein are those of the authors and should not be construed as representing the official views or policies of IARPA, the ODNI or the US Government. Devices were made at the UC Santa Barbara Nanofabrication Facility, a part of the US NSF-funded National Nanotechnology Infrastructure Network, and at the NanoStructures Cleanroom Facility.

**Author contributions** J.K. and R.B. designed the sample and performed the experiment. A.F. and J.M.M. designed the experiment. J.K., R.B., and A.M. fabricated the sample. A.F., J.K., and R.B. analyzed the data. J.K., R.B., A.F., and J.M.M. co-wrote the manuscript. All authors contributed to the fabrication process, experimental set-up and manuscript preparation.

**Author Information** Correspondence and requests for materials should be addressed to J.K. (julian@physics.ucsb.edu) or J.M.M. (martinis@physics.ucsb.edu).

# Supplementary Information for: "State preservation by repetitive error detection in a superconducting quantum circuit"

J. Kelly,[1,*] R. Barends,[1,2,*] A. G. Fowler,[1,3,2,*] A. Megrant,[1,4] E. Jeffrey,[1,2] T. C. White,[1] D. Sank,[1,2] J. Y. Mutus,[1,2] B. Campbell,[1] Yu Chen,[1,2] Z. Chen,[1] B. Chiaro,[1] A. Dunsworth,[1] I.-C. Hoi,[1] C. Neill,[1] P. J. J. O'Malley,[1] C. Quintana,[1] P. Roushan,[1,2] A. Vainsencher,[1] J. Wenner,[1] A. N. Cleland,[1] and John M. Martinis[1,2]

[1]*Department of Physics, University of California, Santa Barbara, CA 93106, USA*
[2]*Present address: Google Inc.*
[3]*Centre for Quantum Computation and Communication Technology, School of Physics, The University of Melbourne, Victoria 3010, Australia*
[4]*Department of Materials, University of California, Santa Barbara, CA 93106, USA*

**CONTENTS**

## I. THE CLASSICAL REPETITION CODE

Suppose we wish to reliably store a single classical bit of information, 0 or 1, for a very long period of time. There are many ways we could attempt to do this. We could write the number on a piece of paper, or carve the number into a boulder, or even scratch the number into a diamond; however, all of these methods are error-prone. Paper burns, ink fades, rocks weather, and diamonds can be stolen. There is no known physical method of truly permanently storing information for later retrieval. Any storage scheme will have some probability of failure per unit time, with the most likely failure mechanism in many schemes being negligent or malicious human activity. We shall quantify all of these failure mechanisms by a single number $p$, the probability of failure per unit time. Without loss of generality, we shall take failure to mean a bit-flip. We can always convert arbitrary errors into bit-flip errors, as lost or unrecognizable data can simply be replaced with a random 0 or 1.

The simplest method of increasing the reliability of information storage is to make multiple copies, and ensure these $n$ copies are subject to errors that are as independent as possible. When using this generic method, we are using a classical repetition code. Multiple paper copies, for example, could be stored in multiple geographic locations. This won't prevent a planet-killing asteroid simultaneously destroying all data, or an organized multi-site human attack, however we shall assume these $n$-bit correlated errors are sufficiently rare to neglect. We shall also assume all $m$-bit, $1 < m < n$ correlated errors are sufficiently rare to neglect and focus on the probability $p$ of each individual bit suffering a flip.

### A. Trusted supervisor

Suppose we have a hypothetical perfectly reliable and trustworthy supervisor at our disposal. Once per unit time, this supervisor could check each bit of data. Each bit has independent probability $p$ of having suffered a bit-flip. If $p$ is small and $n$ is large, most of the data is very likely to still be correct, and the supervisor can take a majority vote, and set the minority bits to the majority value. Note that it is possible that a majority of sites will have suffered an error, and that after "correction" every site will contain the wrong value. The probability of a majority of sites suffering an error is

$$p_{\text{fail}} = \sum_{i=\lceil n/2 \rceil}^{n} \binom{n}{i} p^i (1-p)^{n-i}. \tag{S1}$$

Given the average number of errors is $pn$, and the standard deviation $\sigma = \sqrt{np(1-p)}$, we need $p < 1/2$ to ensure that the average number of errors is less than $n/2$. As $n$ is increased, $pn$ will then be an $O(\sqrt{n})$ increasing number of $\sigma$ below $n/2$, implying exponential suppression of $p_{\text{fail}}$. This example shows how the classical repetition code, given independent errors and a trusted supervisor, can arbitrarily reliably store a single bit of information using only a simple majority vote per unit time. Note that to first order in $p$, $p_{\text{fail}} \sim p^{\lceil n/2 \rceil}$.

### B. Secret data

Suppose now that we wish to keep the data secret. Instead of granting our supervisor permission to look at the data directly, we allow them only to ask if two given bits are the same or not. That is, we allow them to access the result of the exclusive-OR (XOR) operator $\oplus$, where $0 \oplus 0 = 0$, $0 \oplus 1 = 1$, $1 \oplus 0 = 1$, and $1 \oplus 1 = 0$. We shall conceptually arrange our $n$ bits in a line, and focus on the XOR of neighboring bits.

When a single error occurs, away from the ends of the line of bits, the parities of two pairs of bits around the error become 1. Figure S1 gives a detailed worked example showing how this simple case is handled. We represent the parity changes in the graph of Fig. S1d as red colored vertices. The lines between the vertices represent how a single bit error event affects only its two neighboring vertices, except at the ends were the line connects only to an edge vertex. In Fig. S1e we show the obvious decoding of these pair of red vertices into an error chain of a single blue line, which represents the bit-flip. However, the error can also be decoded into the 4 blue lines (error chains) in Fig. S1f, which represents the inverse of the solution in Fig. S1e. As an error chain with $n$ lines has probability $O(p^n)$, assuming independent errors, the solution in Fig. S1e with $O(p)$ is much more likely to occur than for Fig. S1f with $O(p^4)$, and should thus be the choice for the decoding.

Figure S2 gives a more complex example when there are multiple error events. To decode this example, we first show in Fig. S2b the graph of the errors. To decode the multiple red vertices into error events (blue lines), we need to form error chains. A simple algorithm to do this is: (I) From left to right, find the first red vertex and match it with another red vertex to its right. After matching a pair of vertices, continue matching pairs until reaching the right edge. (II) A second solution is the inverse of the above solution. These two solutions are shown in Fig. S2c and d. As d has a shorter total length of the error chains, it is the most likely solution.

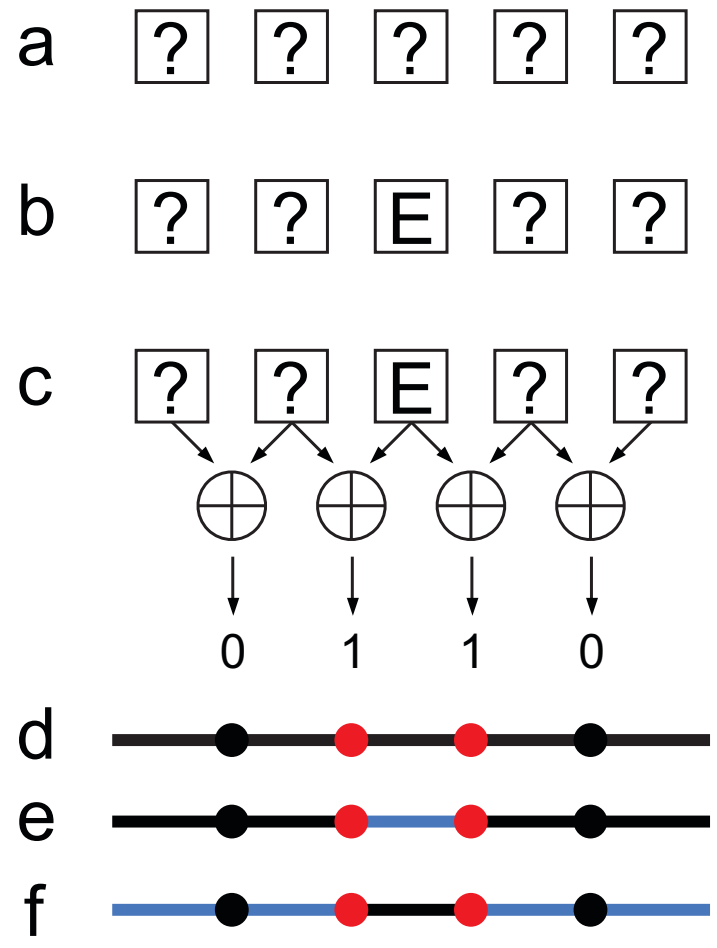

FIG. S1. **Processing perfect parity measurements.** **(a)** Five identical copies of an unknown bit represented by a question mark. **(b)** A bit-flip error on the middle bit, indicated by E. **(c)** Perfect measurement of the parity of neighboring bits would give all 0's in the absence of errors, but in this case two 1's are reported. **(d)** The string of parity measurements can be converted into a graph problem. Each vertex represents one parity measurement, and a red vertex is associated with each error. These 1's are located at the ends of chains of errors. Since errors are independent, error patterns containing fewer errors are more likely. **(e)** The simplest solution is a single error occurring between the red vertices, indicated by a blue line. **(f)** A second but less likely solution is the inverse of the above solution, consisting of two error chains and 4 errors.

The errors are decoded properly if less than $\lceil n/2 \rceil$ errors occur in a single time interval. This means that only having access to parity information is just as powerful in decoding as being able to directly view the data, and we retain $p_{\text{fail}} \sim p^{\lceil n/2 \rceil}$. Indeed, the corrections suggested by matching perfect parity information are identical to the corrections suggested by taking a majority vote given actual data values.

### C. Imperfect parity measurements

When parity measurements are imperfect, we can no longer process each round of parity information independently. For example, if the only error is in a parity measurement and we use the single-round algorithm of the previous section, any graph solution we choose will lead to disastrous corruption of the data, as shown in Fig. S3.

We can fix our decoding algorithm by considering the effect of errors both in space and time. To do so we need to introduce the notion of a detection event, which corresponds to a change of a measurement parity in time. Given arbitrary values on the data bits and assuming no errors, each round of parity measurements will be the same as the previous round. When the parity changes, we know an error must have occurred nearby. For data errors, we see changes as described in the previous section. For parity measurement errors, there is first a change in the parity output, and then in the next cycle a change back to the original (correct) value. Thus a data error produces a pair of detection events in space (with single events at the boundaries), while a parity-measurement error produces a pair of detection events in time. All these errors can be uniquely identified when sufficiently sparse.

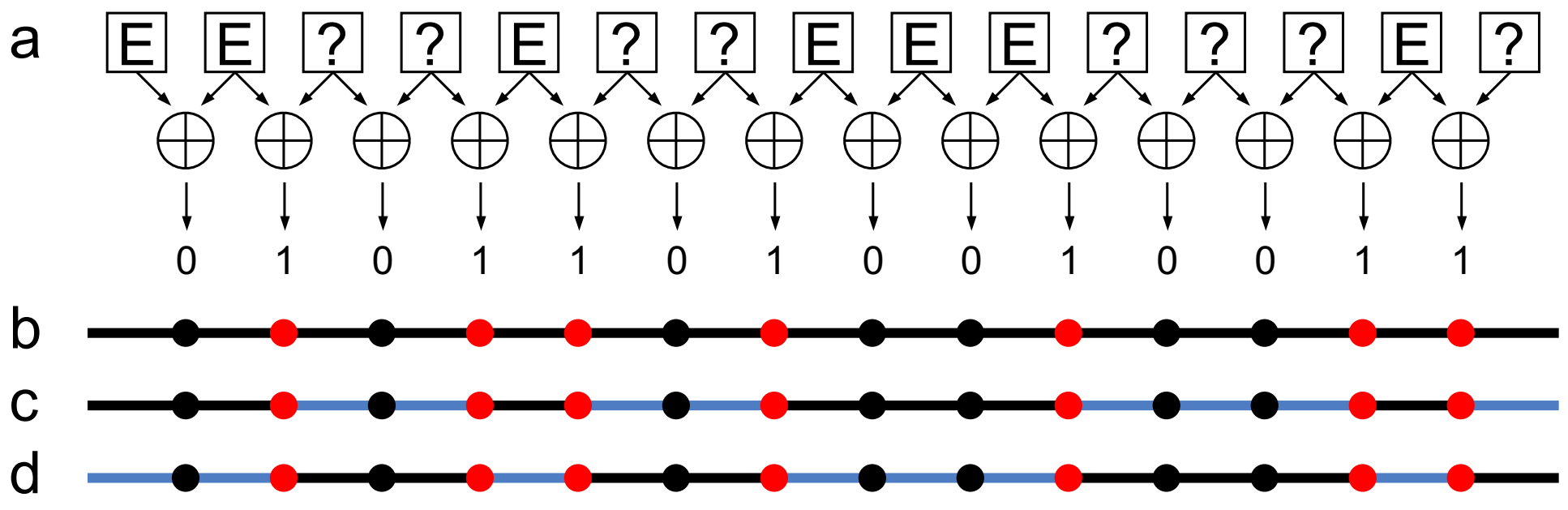

FIG. S2. **Processing perfect parity measurements, large example.** **(a)** 15 initially identical copies of an unknown bit suffer errors at unknown locations leading to a specific pattern of pairwise parity measurements. **(b)** Graph problem corresponding to the parity measurements. Our goal is to connect the colored vertices in pairs or to a graph boundary using the minimum total number of edges. **(c)** Non-optimal weight 8 graph solution. **(d)** Optimal weight 7 graph solution, i.e. contains the fewest errors. Note that after applying corrections corresponding to the blue edges, we will restore the data to its original state using the optimal graph solution, and the bit-inverse of the original state using the non-optimal graph solution. This is a generic property — after correction one will always obtain the original data or its perfect bit-inverse.

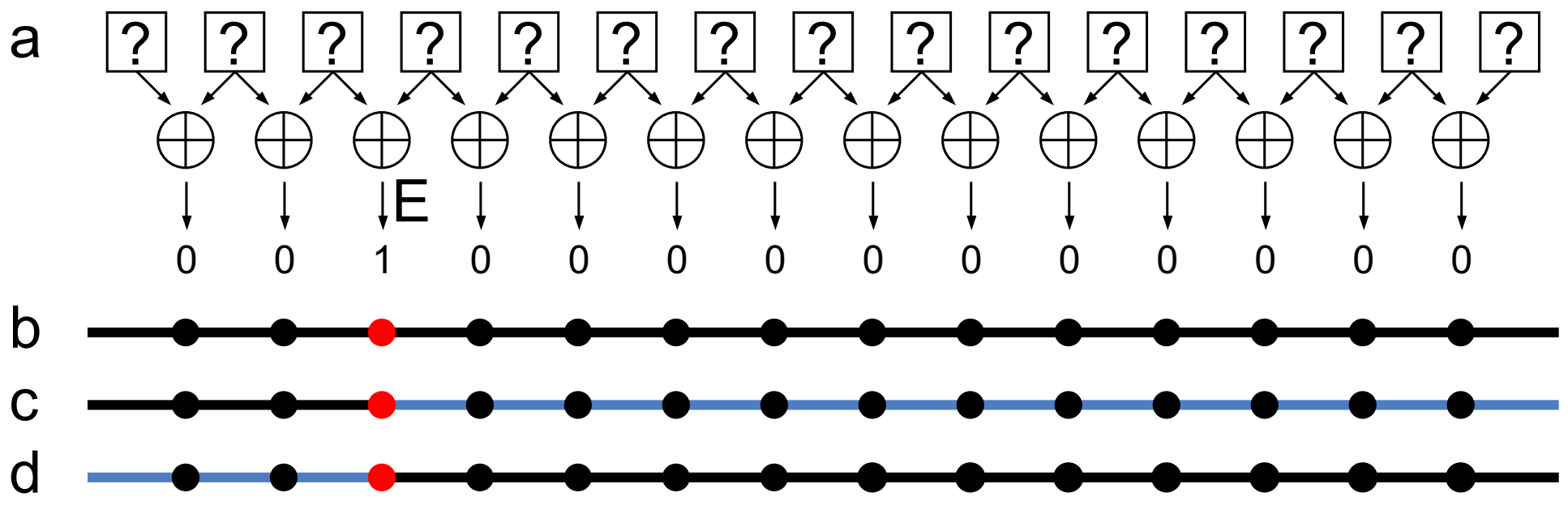

FIG. S3. **Incorrectly processing imperfect parity measurements.** **(a)** 15 initially identical copies of an unknown bit remain free of error, however a single parity measurement reports the wrong result. **(b)** The incorrect parity measurement leads to a single colored vertex. **(c)** Non-optimal weight 12 graph solution corrupting 12 of the data bits. **(d)** Optimal weight 3 graph solution corrupting 3 of the data bits. Note that after applying corrections corresponding to the blue edges, the data will be neither its original state nor its bit-inverse. This implies the described "correction" method is flawed.

We show an example of this behavior in Fig. S4. A data error introduced at $t = 2$ gives a pair of detection events in space, while a measurement error at $t = 4$ gives a pair of detection events at times $t = 4, 5$. This figure also illustrates the basic idea of decoding the detection events into errors using the minimum-weight perfect matching algorithm. Here, a detection event is chosen and a region around it explored until another detection event or edge is found, whereby the two detection events are matched. The idea of the algorithm is that one should connect the red vertices to each other in pairs or to a graph boundary using the minimum total number of lines, where each line corresponds to the location of an error. Importantly, if errors are independent, patterns with fewer lines (errors) are more likely. The number of errors in a pattern is called its weight; an efficient algorithm solving this problem has existed since the mid-60's called minimum-weight perfect matching[1–3]. More details on the algorithm can be found in Ref.[4].

### D. Time boundaries

When the last time boundary is encountered, the algorithm must have additional parity information to correctly match the detection events. As we must directly measure the bits anyway to check whether our decoding efforts were successful, we can use this data itself to compute the final round of parity measurements. As shown in Fig. S5, this allows one to complete the graph, eliminating the future time boundary and enabling all detection events to be processed. Note that although the final data measurements are imperfect, we may model any such errors as coming from a data error in the previous round and treat the final measurements as error free.

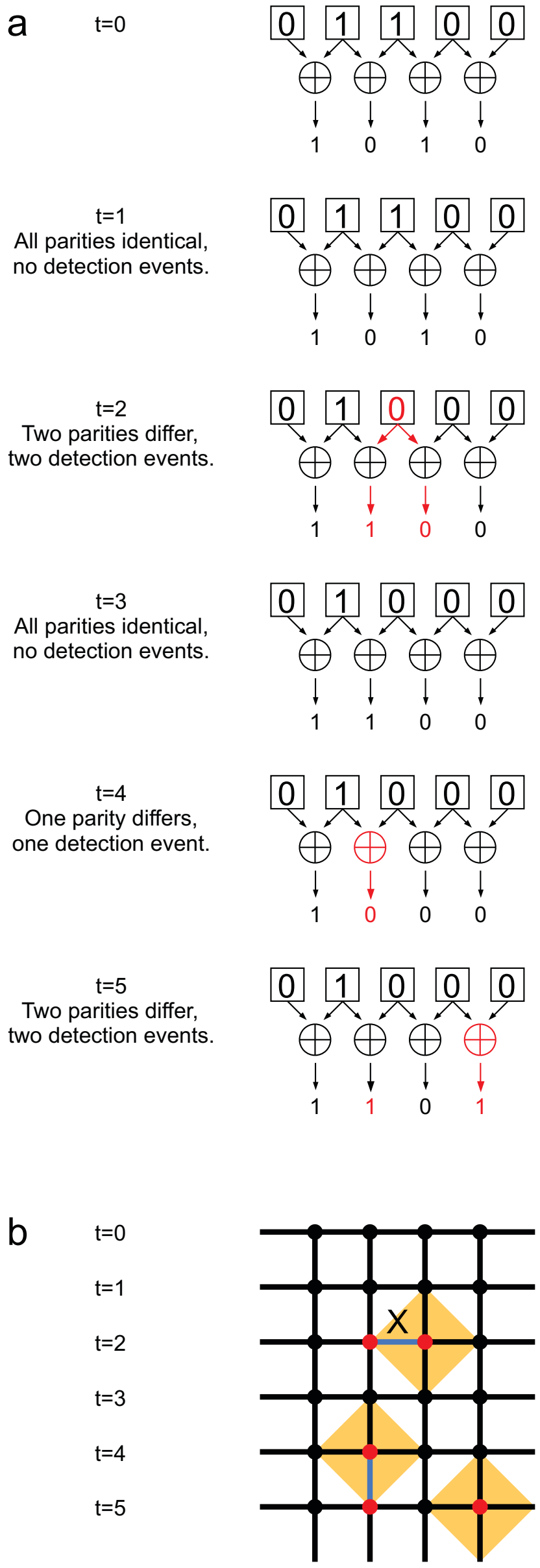

FIG. S4. **Processing imperfect parity measurements.** **(a)** Six rounds of imperfect parity measurements. Every time a parity differs from the previous round, a detection event is generated. **(b)** Corresponding graph problem. Red dots indicate detection events. Note that the graph now includes vertical lines to account for the possibility of parity measurement errors; a single parity measurement error manifests as a pair of sequential detection events. The matching algorithm works by selecting a detection event and exploring uniformly in all directions (orange diamonds) until a feature of interest is encountered. If the feature is another detection event, the two detection events are paired (blue lines). If the feature of interest is the future time boundary, we do not have sufficient information to correctly match the detection event, and need to wait for more data. Corrective data bit-flips are associated with horizontal matched edges. Vertical matched edges indicate parity measurement errors, and require no modification of the data.

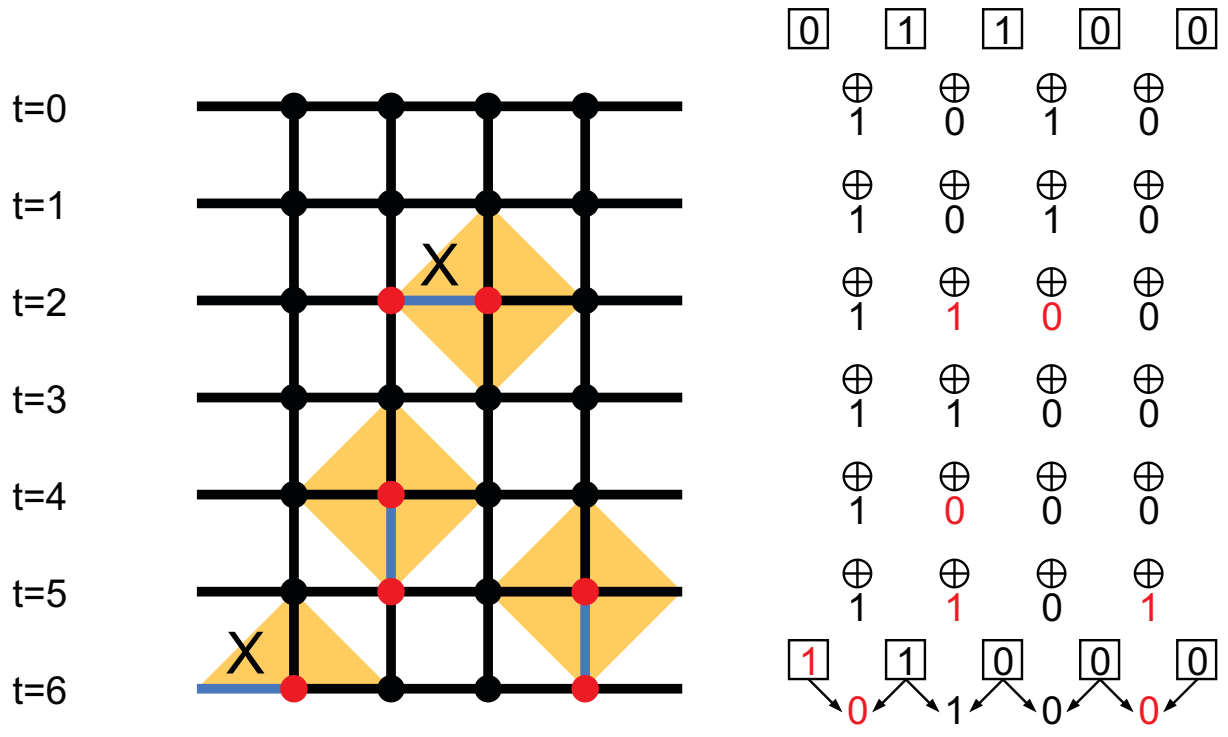

FIG. S5. **Complete processing of imperfect parity measurements.** At $t = 0$, the bits are initialized to 01100 and used to generate the parity measurements. No change at $t = 1$ indicates no initialization errors. At $t = 6$, to check whether storage has been successful, the bits are measured directly (but imperfectly). Any errors during measurement can be treated as errors occurring before a perfect round of measurements, implying no errors remain undetected even when using imperfect physical measurements. This data is used to generate the final parities. In this case, the error pair in the lower right corner is not matched to an edge (a data error), but is correctly identified as a measurement error.

A similar situation occurs for the beginning round, since there is nothing to compare to when computing the parity change at time step $t = 1$. We take the data as perfectly initialized to the desired value, so that errors in initialization are placed as data errors in the first round. The initial parities at $t = 0$ are then computed from this perfect initial state.

### E. Two possible corrected outputs

By moving initialization and measurement errors to detection events, the initial and final states may be considered perfectly known. Logical errors arise only from decoding the detection events. As discussed in the case of only data errors, decoding gives the proper state or its logical error, the bit-inverse. This is also true for the general case of data and measurement errors, as illustrated graphically in Fig. S6. It is interesting to note that the decoding process does not have to match exactly all error events, but only needs to correctly identify the totality of bit errors. As illustrated with the bottom line of Fig. S6d, a logical error only occurs when a net error chain crosses the boundaries, which always produces a bit-inverse of the proper final state.

An advantage of this procedure is that the algorithm removes state preparation and measurement errors (SPAM) to the same order as the error correction itself. This is a hallmark of fault-tolerance in that errors in every part of the quantum circuit are treated equally.

The idea that decoding gives the state or its bit-inverse is perhaps surprising, and although it is a mathematical statement, we have checked our decoding algorithm for consistency. For those who are concerned about using the final mea-surement data to help compute the errors in the final round, we note it is possible to compute errors without this data. In this case one would have greater errors because of imperfect matching of the final detection events, giving errors in both matching and the final measurement. As these additional errors come from the measurement, they are constant with changing the number of rounds, and thus the decay of the fidelity will be the same as for matching with the full measurement data.

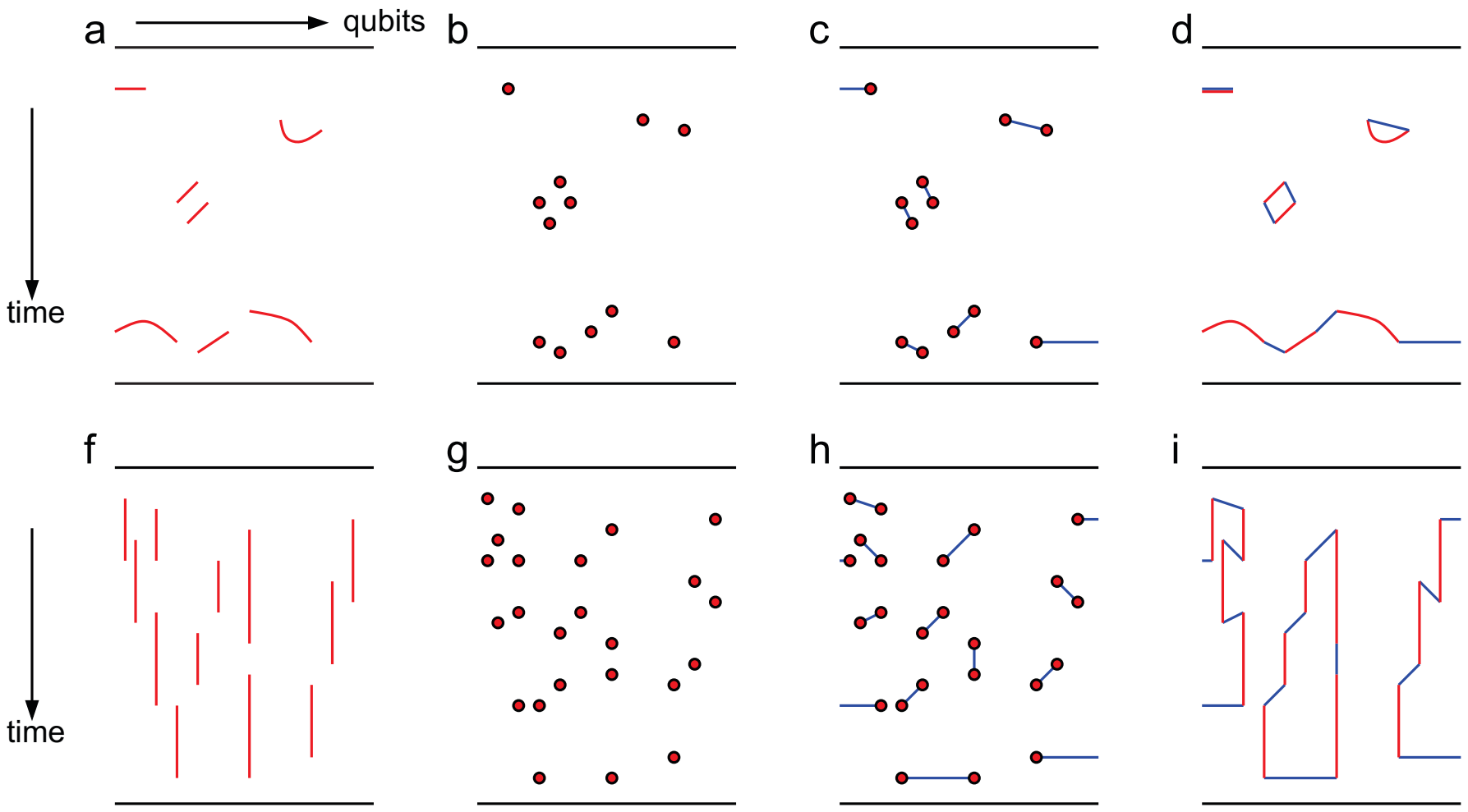

FIG. S6. **Topology of corrections.** **(a)** A large repetition code running for many cycles with error chains marked in red. **(b)** Corresponding detection events. **(c)** A minimum-weight matching of the detection events. **(d)** Errors and corrections plotted together. Successful corrections together with their associated errors form rings, or U-shapes starting and ending on a single boundary. Unsuccessful corrections together with their associated errors form chains connecting different boundaries. Given every horizontal edge corresponds to either an error or a correction, rings and U-shapes contribute a net identity operation. Chains connecting different boundaries contribute a net single bit-flip on every bit. Note that this is true even if the chain doubles back on itself, as then some bits will get an odd number of flips that will cancel down to a net single bit-flip. This is why, even in a physical system with realistic errors, after correction the output will either match the input or be its exact bit-inverse. **(e–i)** A more complex example focussing on extreme measurement error. In this case, errors and corrections cancel leading to successful storage.

## II. QUANTUM REPETITION CODE: THEORY

Given qubits instead of bits, we need to be able to protect quantum superpositions from error. Qubits encode information in amplitude and phase, which can be expressed in terms of $\hat{X}$ and $\hat{Z}$ operators. Thus, errors can be expressed in terms of bit-flip ($\hat{X}$) and phase-flip ($\hat{Z}$) errors. However, detecting both types of errors simultaneously is nontrivial, as $[\hat{X},\hat{Z}] \neq 0$. This can be overcome by constructing operators that measure the parity of two qubits. Take for example, the $\hat{X}_1\hat{X}_2$ and $\hat{Z}_1\hat{Z}_2$ operators:

$$\begin{aligned}
[\hat{X}_1\hat{X}_2, \hat{Z}_1\hat{Z}_2] &= (\hat{X}_1\hat{X}_2)(\hat{Z}_1\hat{Z}_2) - (\hat{Z}_1\hat{Z}_2)(\hat{X}_1\hat{X}_2) \\
&= \hat{X}_1\hat{Z}_1\hat{X}_2\hat{Z}_2 - \hat{Z}_1\hat{X}_1\hat{Z}_2\hat{X}_2 \\
&= (-\hat{Z}_1\hat{X}_1)(-\hat{Z}_2\hat{X}_2) - \hat{Z}_1\hat{X}_1\hat{Z}_2\hat{X}_2 \\
&= 0
\end{aligned} \tag{S2}$$

Thus, these multi-qubit operators can be used to detect the bit and phase parity of two qubits without knowing and collapsing the individual state of each qubit – analogous to the classical secret information example.

Consider the operator $\hat{Z}\hat{Z}$. This operator has the property that $\hat{Z}\hat{Z}|00\rangle = +|00\rangle$, $\hat{Z}\hat{Z}|01\rangle = -|01\rangle$, $\hat{Z}\hat{Z}|10\rangle = -|10\rangle$, and $\hat{Z}\hat{Z}|11\rangle = +|11\rangle$. This operator can detect changes in parity of the qubits; however, it cannot determine which qubit has flipped. To overcome this, we use a one-dimensional array of qubits and nearest neighbour parity operators, similar to the array of bits and XORs in the classical example. As we must protect qubits from both $\hat{X}$ and $\hat{Z}$ errors simultaneously, a fully protected state requires a two-dimensional array of qubits. Here, we design our experiment to focus on a chain of qubits and $\hat{X}$ errors only, as this is experimentally viable today.

We can construct the $\hat{Z}\hat{Z}$ operator through the use of quantum logic gates and measurement, as seen in Fig. S7. Each controlled-NOT (CNOT) gate will flip the top ancilla qubit dependent on the state of the control qubit, just like the classical XOR operation. Thus, the state $|00\rangle$ will map the ancilla qubit to $|0\rangle$, the $|01\rangle$ or $|10\rangle$ states will map the ancilla to $|1\rangle$, and the $|11\rangle$ will have two flips that cancel, mapping the ancilla to $|0\rangle$. After the circuit is executed, the measured state of the ancilla qubit will encode the eigenvalue of the $\hat{Z}\hat{Z}$ operator.

## III. QUANTUM REPETITION CODE: EXPERIMENT

Our experimental implementation of the quantum repetition code consists of nine qubits total, five data qubits and four measure qubits. Limited memory in our control electronics restricts us to a maximum of eight parity measurement cycles. One complete run of our experiment consists of qubit initialization, between one and eight cycles of the physical gate sequence in Fig. S26, data qubit measurement, then postprocessing to determine whether a logical error has occurred. This section focuses on the postprocessing, with emphasis on doing this with care to achieve the lowest possible logical error rate.

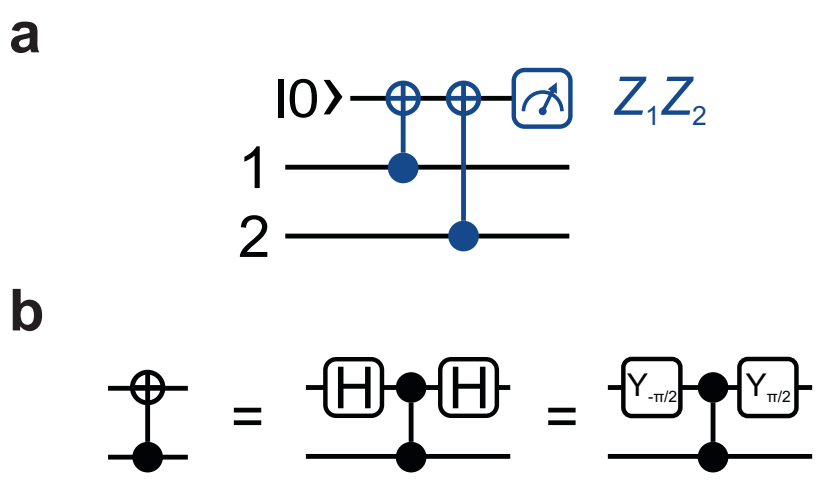

FIG. S7. **Parity measurement circuit.** (a) CNOT gates flip the state of the top ancilla qubit conditional on the state of the control qubits. The two CNOT gates and measurement will encode the eigenvalue of the $\hat{Z}_1\hat{Z}_2$ operator onto the state of the ancilla qubit, see text. (b) The CNOT gate can be decomposed into a controlled-phase (CZ) gate plus single-qubit gates. Decomposition can either consist of a Hadamard on the target qubit before and after the CZ, or a $Y_{-\pi/2}$ before and $Y_{\pi/2}$ after on the target qubit.

Any information pertaining to the physical performance of the device can be incorporated into the postprocessing to ensure that the very best possible corrections are suggested at the end of a run. We shall discuss five increasing levels of detail, with the final level of detail corresponding to the results reported in the main text. We do not claim to have exhausted available techniques.

### A. Basic processing

There are a number of steps required to identify physical errors given the raw experimental output. Example data moved through each of these steps is shown in Fig. S8. In the following sections, we will explain in detail how each of these steps is performed.

(a) **Raw Data.** Before we can begin discussing more advanced processing, we must give a little more detail on how the experimental output is converted into detection events. Our experiment makes use of QND measurement and does not reinitialize measure qubits to $|0\rangle$ as was done in Fig. S7a. This makes detection event identification more complex.

An example of experimental data gathered during an eight cycle run is shown in Fig. S8a. Some postprocessing has already occurred, namely the conversion of the measurement microwave waveform into a best guess of the corresponding state. This postprocessing is described elsewhere (Section XI). The first line "in" shows which state the nine qubits were intended to be initialized to. The five data qubits have been initialized to $|0_L\rangle$. The eight numbered lines show the output of each measurement qubit for each cycle, and the final line contains the experiment terminating data qubit measurements.

(b) **Additional simulated rounds.** In order for the first cycle of measurements to look for a change in parity, we generate two artificial rounds from a parity computation of the desired initial state. See Fig. S8b. As the data qubits are initialized to either all 0's ($|0_L\rangle$) or all 1's ($|1_L\rangle$), the computed parity is all 0's. Likewise, an additional parity round is computed from the measurement of the final data.

(c) **Calculating detection events.** Consider the action of the parity measurement circuit in Fig. S7a. If just one of the data qubits is in state $|1\rangle$, the value of the measure qubit will be flipped. If the data qubits have the same value, the measure qubit will be unchanged. In the absence of errors, QND measurement and reinitialization therefore leads to two possible behaviors — alternating and constant, see main text. The presence of an error is indicated by a change between alternating and constant behavior. For example, the sequence of measurement results 011 shows a change from alternating to constant behavior, and hence is associated with a detection event.

We desire a simple formula to identify detection events. Given a sequential pair of measurement results, $m_{t-1}$, $m_t$, we can use XOR to detect alternating $m_{t-1} \oplus m_t = 1$ and constant $m_{t-1} \oplus m_t = 0$ behavior. Given three sequential measurements, $m_{t-2}$, $m_{t-1}$, $m_t$, we can detect a change between alternating and constant behavior using $(m_{t-2} \oplus m_{t-1}) \oplus (m_{t-1} \oplus m_t) = m_{t-2} \oplus m_t$. In our experiment, a detection event is generated when a given parity measurement differs from the value reported two rounds ago.

The data converted into detection events is shown in Fig. S8c. Our goal is to use the detection events to calculate bit-flips to apply to the final measurement results, with a success being recorded when the corrected output matches the input.

(d) **Detection events on error connectivity graph.** If we take no details of the underlying quantum circuit into account, and assume here for simplicity that we can only have parity measurement errors or data qubit errors between cycles of the repetition code, we can perform the classical postprocessing exactly as described in Subsection I C for the classical repetition code with imperfect parity measurements. This means using a graph with a square structure, with vertices at every potential detection event location. Vertices associated with actual detection events get colored red (Fig. S8d). A potential data qubit error between a particular pair of repetition code cycles can be visualized as a horizontal edge, a potential measurement error as a vertical edge.

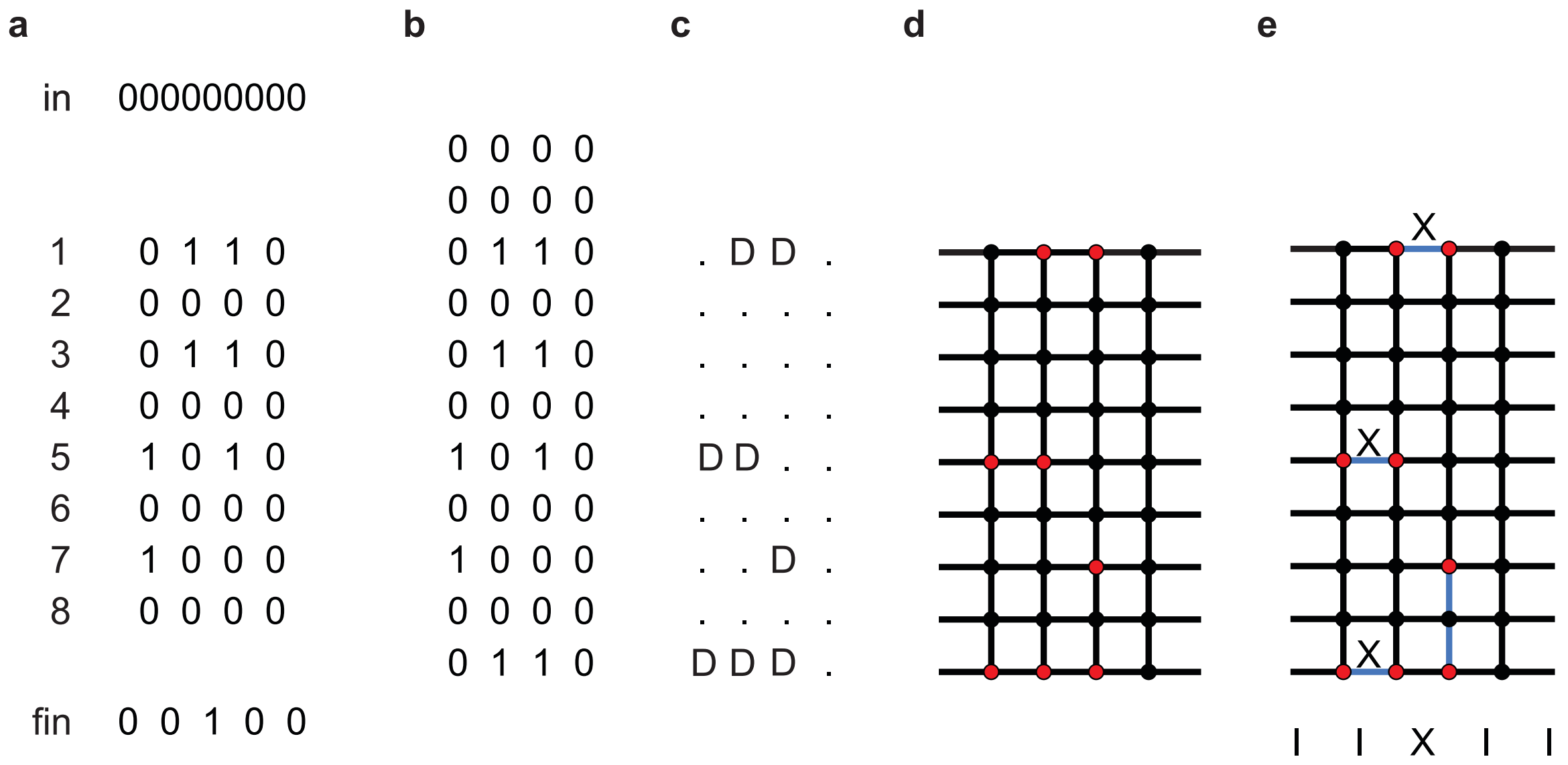

FIG. S8. **Flow of experimental data.** (a) Raw experimental data. (b) Raw experimental data appended by computed data to allow the first round of measurements to detect errors, and a final round of measure qubit outcomes inferred from the data qubit outcomes. (c) Data converted to detection events, indicated by D. (d) Detection events (red circles) placed onto graph which contains information about how errors propagate. (e) Detection events matched together (blue lines) to identify physical errors. The final two detection events on the third measure qubit, connected over three rounds, are an error chain. Final correction to recover data qubit input is shown at the bottom of the graph: $\hat{I} \otimes \hat{I} \otimes \hat{X} \otimes \hat{I} \otimes \hat{I}$.

(e) **Identifying errors.** As before, we use minimum-weight perfect matching to find a minimal set of errors reproducing the observed detection events. Blue edges show that, in this case, a unique minimum set of edges solves this problem. In general, the problem can have multiple solutions. Our processing deterministically but arbitrarily chooses one of these multiple solutions. In this instance, the correction suggested by the solution restores the observed output to the input, and the run has been successful. For the data set used for this paper, 3.35% of 9-qubit 8-cycle runs failed using this method of postprocessing.

For the remainder of this document, experimental data is organized as in Fig. S8.

### B. Data errors during the repetition code cycle

In the previous subsection, our processing assumed that data qubit errors could only occur between repetition code cycles. Despite the fact that this assumption is not, in fact, true, we gave a nontrivial example where the postprocessing succeeded. In reality, data qubit errors can occur at any time. Referring to Fig. 2 of the main text, an error on data qubit 2 during the third 20 ns window will be detected on measure qubit 3 immediately, but not on measure qubit 1 until the next cycle. Similarly, an error on data qubit 4 during the second 20 ns window will be detected on measure qubit 5 immediately, but not on measure qubit 3 until the next cycle. Finally, an error on data qubit 6 during the second or third 20 ns windows will be detected in the same cycle on measure qubit 5, but not on measure qubit 7 until the next cycle. Taking into account these three new classes of error means including three new classes of edges in our graph problem.

Consider the following run.

| | | |
|---|---|---|
| in | 101010101 | |
| | 0 0 0 0 | |
| | 0 0 0 0 | |
| 1 | 0 0 0 0 | . . . . |
| 2 | 0 0 0 0 | . . . . |
| 3 | 0 0 0 0 | . . . . |
| 4 | 0 0 0 0 | . . . . |
| 5 | 0 0 0 0 | . . . . |
| 6 | 1 0 1 0 | D . D . |
| 7 | 0 1 0 1 | . D . D |
| 8 | 1 1 1 0 | . D . . |
| fin | 0 0 1 0 1 | D D . . |

Example 1. An example run (real data) illustrating the benefit of considering data qubit errors during the repetition code cycle.

Using the basic postprocessing of Subsection III A, the graph, solution, and suggested corrections are shown in Fig. S9b, leading to failure. By including additional edges corresponding to data qubit errors during the repetition code cycle, a unique and better solution becomes available, resulting in a successful run. The 9-qubit 8-cycle logical error rate drops from 3.35% to 3.29% after including the additional edges. The improvement is small, as expected, as it is rare that errors occur in the precise windows required to generate behavior handled by these extra edges.

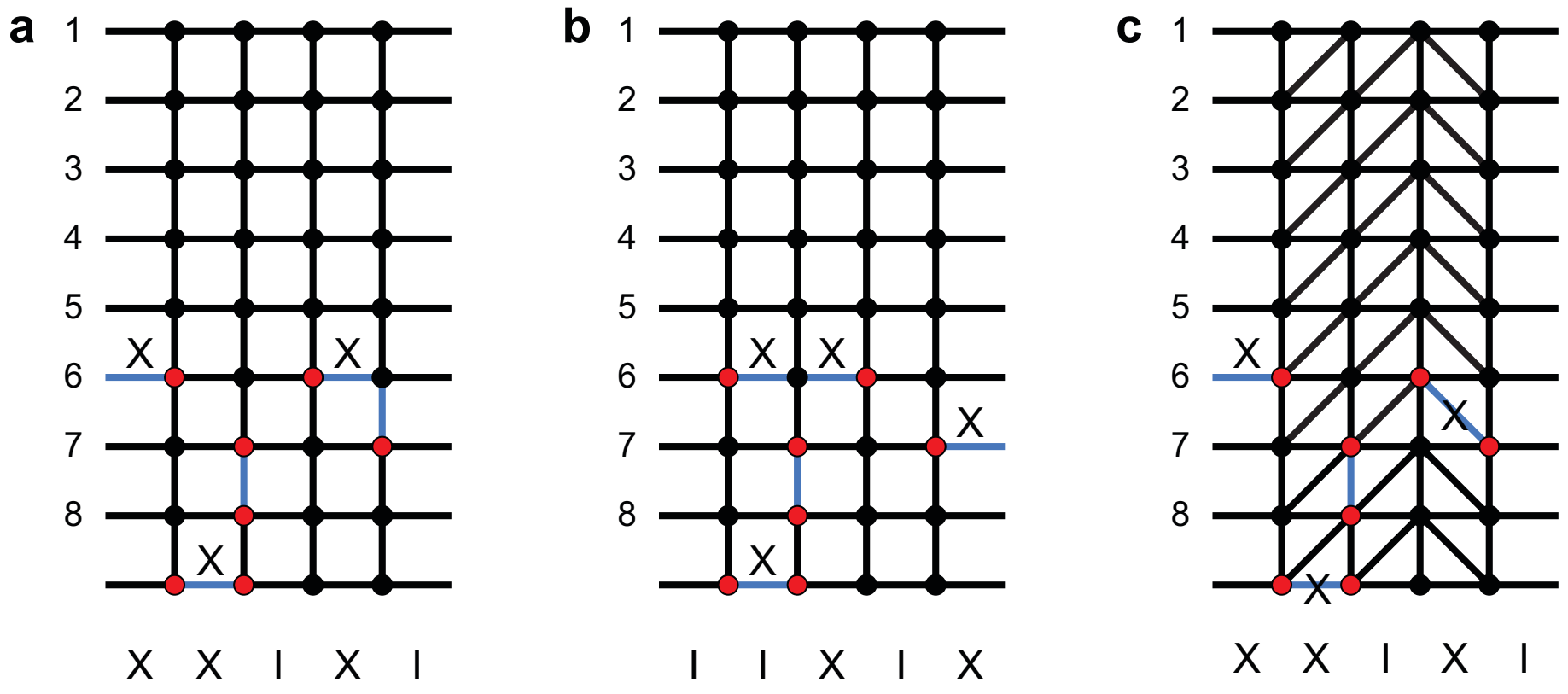

FIG. S9. Graph problems corresponding to Ex. 1. Parts **(a)** and **(b)** show two possible solutions when assuming only basic errors as was done in Subsection III A. Our code deterministically but arbitrarily chooses solution **(b)**, which for this experimental run was the wrong thing to do, resulting in failure. Without any additional information, solutions **(a)** and **(b)** both use 5 edges and are equally good. In part **(c)**, additional edges have been added to account for the possibility of data qubit errors during the repetition code cycle. When this is done, a unique 4 edge solution is found by postprocessing, successfully restoring the output to the input. Fewer edges means fewer errors and hence a more likely error pattern that matching should favor.

TABLE S1. Input error model.

| Gate | Error |
|---|---|
| CZ | 1% |
| X | 0.1% |
| Idle (20 ns) | 0.05% |
| Initialization | 2.5% |
| Readout (measure qubit) | 1.5% |
| Readout (data qubit) | 3%. |

### C. Weighted edges

In the previous Subsection, it was noted that there are few errors that can lead to a diagonally offset pair of detection events. It would seem reasonable, therefore, to make it less likely for postprocessing to choose diagonal edges when multiple options are available. Calculating the probability of each edge first requires an error model for every gate in the repetition code cycle. As a first pass, we choose to believe that every gate suffers errors well modeled by a depolarizing channel, and that gates of the same type suffer errors at the same rate.

We input the error rate associated with each operation in table S1, determined from previous techniques[5,6]. These operations are all that are required during our experiment. By studying exactly where and when every possible error is detected, we can determine the probability of every edge in our graph. We convert each edge probability $p_i$ into a weight $w_i = -\ln p_i$ so that the weight of two consecutive edges is $w_i + w_j = -\ln p_i - \ln p_j = -\ln p_i p_j$. This ensures that minimum weight perfect matching[1–3] will consider two hypothetical detection events that can be matched either through independent single edges to a nearby boundary or to each other through two links to have two equally good matchings, as we wish, if the edge probabilities are equal.

Consider the following run.

```
in   101010101
      0 0 0 0
      0 0 0 0
1     0 0 1 0      . . D .
2     0 0 1 0      . . D .
3     0 0 0 0      . . D .
4     0 0 0 0      . . D .
5     0 1 0 0      . D . .
6     0 0 0 0      . . . .
7     0 1 0 0      . . . .
8     1 1 0 0      D D . .
fin  0 1 0 1 1     . D D .
```

Example 2. An example run (real data) illustrating the benefit of carefully weighting each edge in the graph to reflect the actual probability of detection events being observed at the edge's vertices. Diagonal edges, in particular, are much less likely than horizontal and vertical edges.

Without taking the error rate of each gate into account, the corresponding graph problem has equally weighted edges and hence two equally acceptable solutions (Fig. S10), only one of which leads to successful correction. With weights set according to their probability ($w = -\ln p$), low probability diagonal edges acquire high weights and the solution in Fig. S10b, in this case the correct solution, becomes favored resulting in a successful run. The 9-qubit 8-cycle logical error rate drops from 3.29% to 2.897% through the use of weighted edges, a significant improvement. This is expected as a great deal of physical information has now been added to the graph, helping the postprocessor accurately distinguish between many previously degenerate solutions.

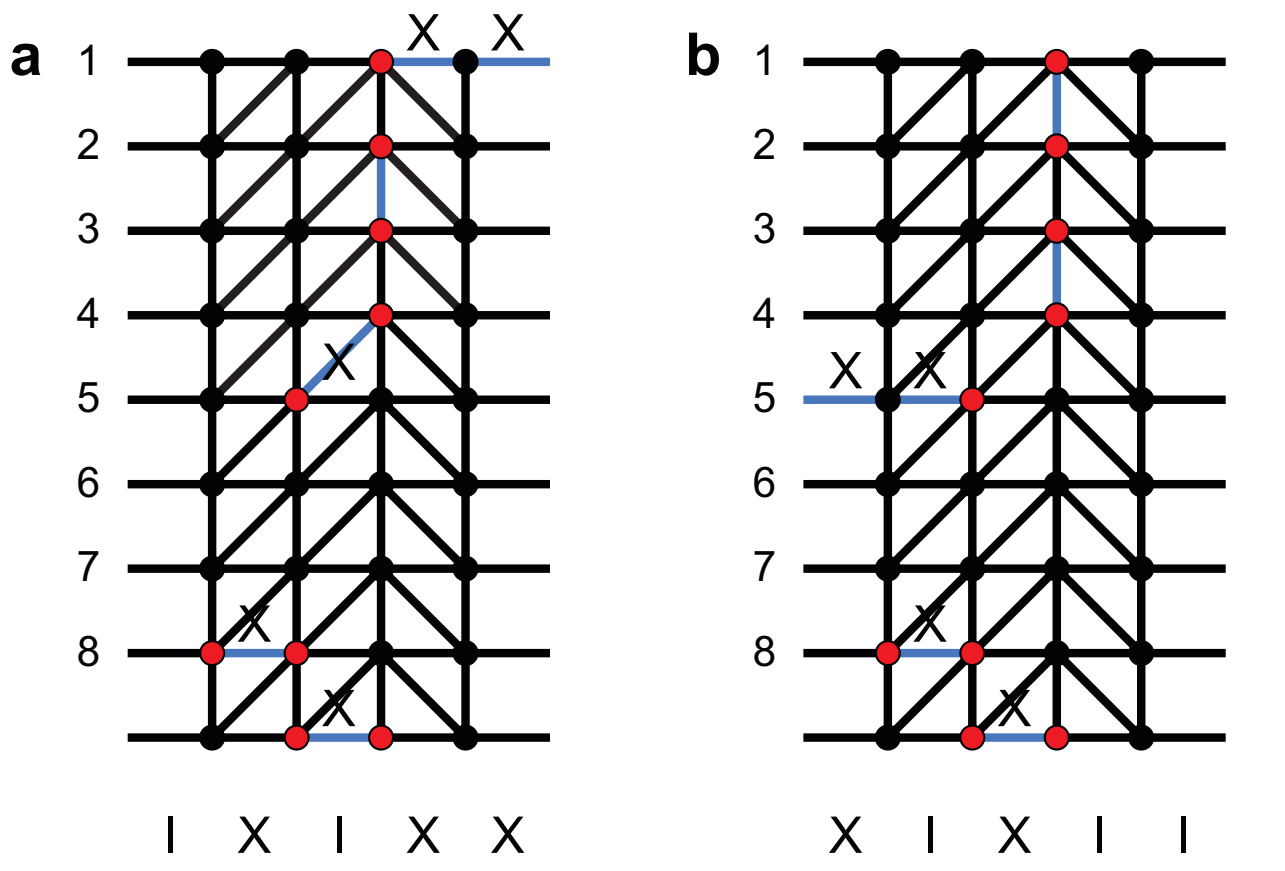

FIG. S10. Graph problem corresponding to Ex. 2. With equally weighted edges, there are two equally good solutions with six edges. Our software deterministically chooses solution **(a)**, which in this case does not restore the output to the input, resulting in a failed run. By weighting edges according to their probability, diagonal edges get a much higher weight as they are low probability ($w = -\ln p$), penalizing their use. This breaks the degeneracy between these two solutions, favoring solution **(b)** and leading to a successful run.

## D. Leakage

Superconducting qubits are not two-state systems, and leakage to other non-computational states can and does occur. When a measure qubit outputs 2, which will be the input to the next repetition code cycle, the next measurement result will be unreliable. We can only accurately predict what the next measurement result should be, even in the absence of errors, if the cycle begins with 0 or 1 on the measure qubit. This physical understanding can be fed into the postprocessing by setting the probability of a vertical edge to be 0.5 conditional on observing 2 at the end of the first cycle it is connected to.

Furthermore, a $|2\rangle$ on a measure qubit at the beginning of the repetition code cycle can induce errors on the neighboring data qubits. We observed that a neighboring data qubit in a $|1\rangle$ state is essentially randomized, whereas a neighboring data qubit $|0\rangle$ is mostly undisturbed. We can model this by setting the probability of the two horizontal edges associated with the next repetition code cycle to 0.25. These two reweighting rules are depicted in Fig. S11

Consider the following run.

| | | |
|---|---|---|
| in | 101010101 | |
| | 0 0 0 0 | |
| | 0 0 0 0 | |
| 1 | 0 0 0 0 | . . . . |
| 2 | 0 0 0 0 | . . . . |
| 3 | 0 2 1 0 | . D D . |
| 4 | 1 1 1 0 | D D D . |
| 5 | 0 1 2 0 | . . . . |
| 6 | 1 0 2 0 | . D . . |
| 7 | 0 2 2 2 | . . . D |
| 8 | 1 1 2 2 | . D . D |
| fin | 0 0 0 0 1 | D . . D |

Example 3. An example run (real data) illustrating the benefit of lowering the weights of future edges after a measure qubit reports 2. Note that a 2 is treated as a 1, plus reweighting.

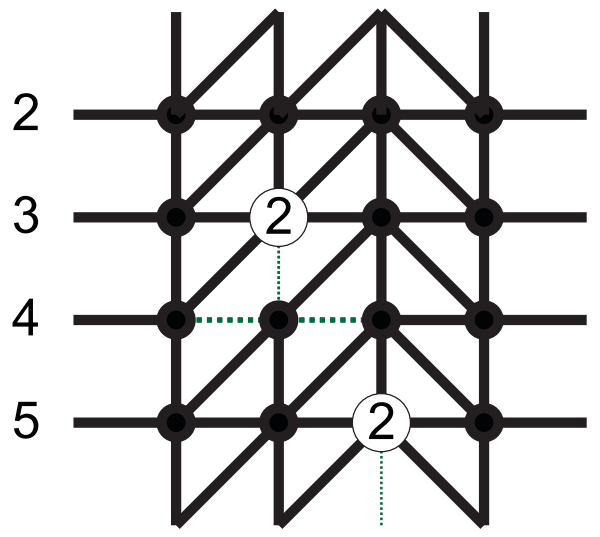

FIG. S11. When a measure qubit reports 2, the next measurement will be unreliable, and this knowledge can be accounted for by setting the probability of error for the next vertical edge to be 0.5, resulting in a very low weight link (narrow vertical line). Note that thinner lines are less weight, and therefore more likely. Neighboring data qubits, if they are in the $|1\rangle$, were observed to be strongly corrupted, which can be accounted for by setting the horizontal edges of the next cycle to have probability of error of 0.25 (slightly less narrow horizontal lines). These three high probability (low weight) edges will be strongly preferred by the matching algorithm, more reliably pairing up cascades of detection events resulting from leakage.

A large region of the data has been corrupted by leakage. Without accounting for leakage, the matching shown in Fig. S12a is found, and the corresponding corrections are unsuccessful. With edge reweighting, detection events arising from leakage are preferentially matched to each other, in this instance leading to successful correction. Leakage plays a significant role in superconducting circuits. Including $|2\rangle$ state intelligence in the measurement and postprocessing reduces the 9-qubit 8-cycle logical error rate from 2.897% to 2.414%, a significant improvement.

## E. Fine-tuning the error model

In Subsection III C, we chose simple depolarizing error rates for each type of gate and assumed that every gate of the same type had the same error rate. We shall now relax that assumption, assuming only that every gate of the same type applied to the same qubits will have the same error rate.

To begin, we associate a variable with each geometrically- or weight-distinct edge. The set of variables is represented graphically in Fig. S13. Geometrically-distinct means connecting different sets of qubits, or being horizontal versus diagonal. Weight-distinct means that while the edges may be geometrically-identical, they have different weight. The final row of horizontal edges has a different weight to every other row as the final 3% error data qubit measurements contribute to the generation of detection events associated with this row only, lowering their weights below that of other horizontal lines. Similarly the 2.5% initialization error contributes to the first row of horizontal and vertical edges only. The remaining green edges control the rest of the graph, with a small number of variables controlling the repetitive weight structure.

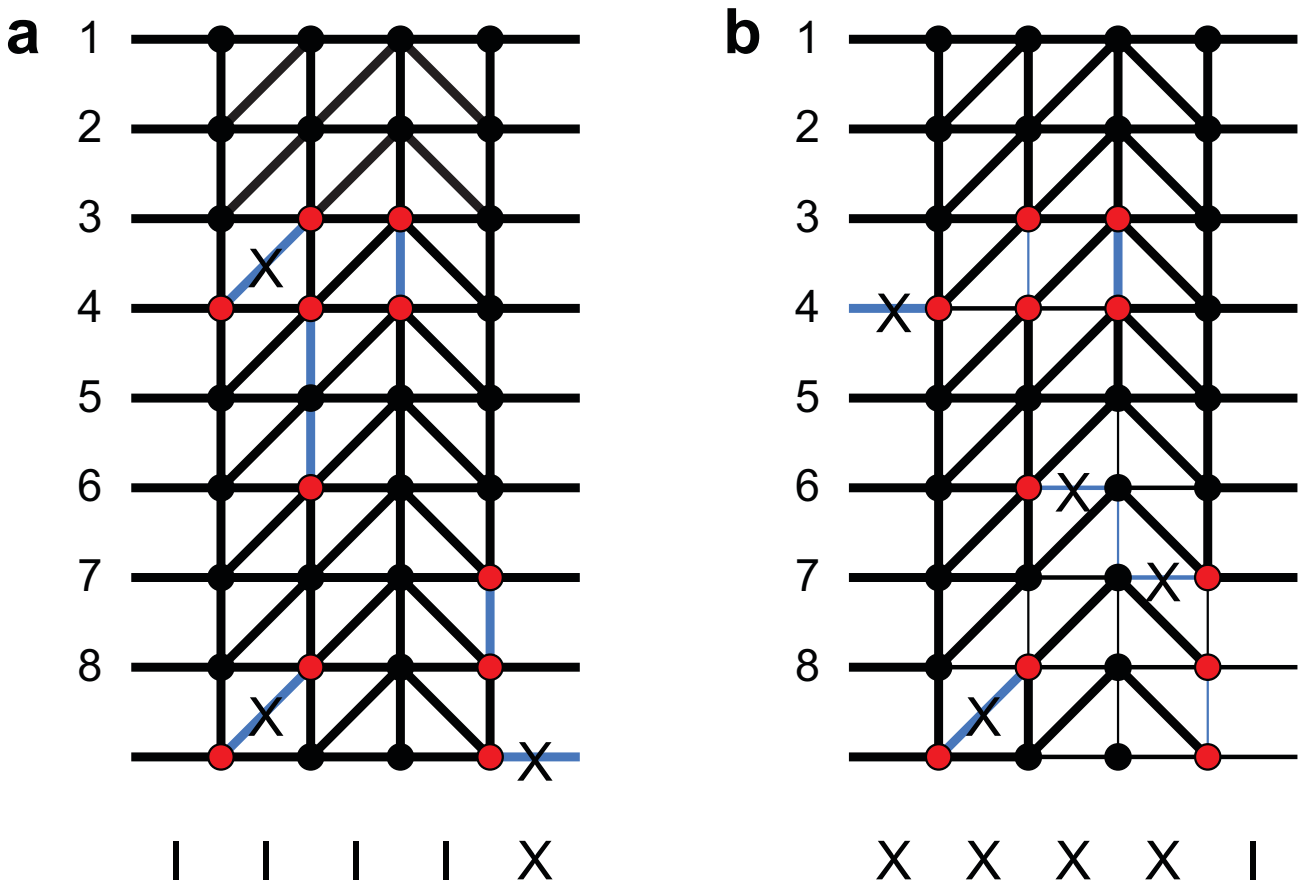

FIG. S12. **(a)** matching of Ex. 3 without accounting for leakage. **(b)** leakage is well handled by lowering the weight of nearby edges following detected leakage. These low weight edges encourage detection events generated as a result of leakage to be paired together leading to a higher probability of successful correction. In this example, after reweighting the suggested corrections successfully return the output to the input.

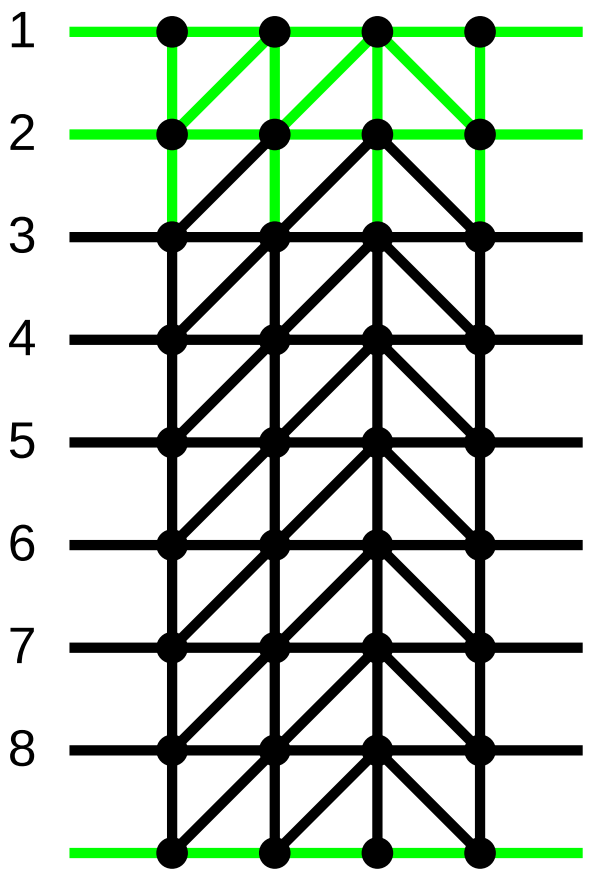

FIG. S13. Green edges denote geometrically- or weight-distinct edges. Every black edge is equivalent to a green edge by time translation. The relatively low number of unique edges is due to the cyclic nature of the repetition code, and hence the cyclic nature of error propagation. Two edges for coping with leakage are also optimized, but not shown as they are dynamically added when leakage is detected.

Our primary data file contains 90,000 runs for each possible number of cycles from 1 to 8 and both $|0_L\rangle$ and $|1_L\rangle$ as input, for a total of 1.44 million runs. When our postprocessing is run on this file using the initial depolarizing error model, 18,750 runs lead to logical errors. We can use the total number of logical errors as a fitness metric and optimize the edge variables defined above for a total logical error count. We make use of the standard Nelder-Mead algorithm to perform this optimization. Optimizing the error model in this manner reduces the 9-qubit 8-cycle logical error rate from 2.414% to 2.246%.

Optimizing using the entire data set is cheating, however, since it doesn't guarantee good performance on unseen data. To combat this, we first split the full dataset in two parts. The chosen split was between even and odd entries in the dataset, which ensures full sampling over the time of the experiment and input states. Second, we use these two sets to train two error models. We obtain logical error rates of 2.222% for the even, and 2.254% for the odd dataset. Note that the logical error rates differ only in the third significant figure, illustrating that the optimized performance is not strongly dependent on the dataset.

Third, to further test the validity of these error models, we test their performance on the datasets they have not seen. We obtain logical error rates of 2.300% for the even error model on the odd dataset and 2.243% vice versa. We note that the performance here is very close to the full dataset, and that indeed, the error model trained on odd data, performs even slightly better when used on the even dataset. These results underline that the error models are nearly equivalent, and therefore can be used to infer a reliable logical error rate. We took the worst results, namely optimizing on even and running on odd data, to be the foundation of the discussion in the main Letter.

TABLE S2. Summary of logical error rates.

| algorithm | logical error rate |
|---|---:|
| basic errors only | 3.350 % |
| data errors during the cycle | 3.290 % |
| weighted edges | 2.897 % |
| leakage | 2.414 % |
| fine-tuning the error model | 2.300 % |

In summary, our most basic postprocessing led to a logical error rate of 3.35%, and our final hardware-optimized postprocessing reduced this to 2.3%, a 30% reduction. Our stepwise optimization are summarised in table S2. In the next section, we shall see that there is a better figure of merit characterizing the performance of a fault-tolerant quantum error correction system. We think that this is a significant test of the theory that better error models give a lower logical error rate.

## IV. ERROR SUPPRESSION FACTOR $\Lambda$ - FIGURE OF MERIT FOR FAULT-TOLERANT QUANTUM ERROR CORRECTION

This Section discusses a universal figure of merit characterizing the performance of any fault-tolerant topological quantum error correction system, meaning the totality of both the quantum hardware and the classical postprocessing. Clearly, the 9-qubit 8-cycle logical error rate we discussed in the previous Section is highly specific to our current experiment, and unlikely to be appropriate for comparison with other hardware or experiments. We seek instead a generic figure of merit characterizing how rapidly errors are suppressed as qubits are added to the system. This means studying and comparing the performance of different orders of fault-tolerance.

In the main text we defined $n$-th order fault-tolerance to mean a system guaranteeing correction of any combination of $n$ errors. Our full 9-qubit experiment is 2nd-order fault-tolerant to $X$ errors. There are three possible 5-qubit subsets that are 1st-order fault-tolerant. In an effort to make the most reliable extrapolations to higher orders, instead of separately running the three possible 5-qubit subsets, we infer the performance of these subsets directly from the full 9-qubit data. This simply means discarding appropriate columns from a larger 9-qubit run. We believe this will give a more accurate estimation of $\Lambda$ than running three separate 5-qubit experiments, which is experimentally taxing.

For example, the performance of the first 5 qubits can be inferred from 9-qubit data as follows.

| | | |
|---|---|---|
| in | 000000000 | 00000 |
| 1 | 0 1 1 0 | 0 1 |
| 2 | 0 0 0 0 | 0 0 |
| 3 | 0 1 1 0 | 0 1 |
| 4 | 0 0 0 0 | 0 0 |
| 5 | 1 0 2 0 | 1 0 |
| 6 | 0 0 0 0 | 0 0 |
| 7 | 1 0 0 0 | 1 0 |
| 8 | 0 0 0 0 | 0 0 |
| fin | 0 0 1 0 0 | 0 0 1 |

Example 4. Left, full 9-qubit run data. Right, the same data restricted to the first five qubits.

This works because the repetition code is topological – its structure is transversely invariant as it grows and its classical processing is local on average. The extra piece of unused repetition code in the above example fundamentally just interacts with the rightmost data qubit introducing a small amount of additional error on this qubit. This in no way conceptually changes the performance or required postprocessing of the leftmost five qubits.

Ideally, a well-constructed topological quantum error correction system should have a logical error rate that scales as $\epsilon \sim 1/\Lambda^{n+1}$. The universal figure of merit $\Lambda$ specifies how much the logical error rate will go down if the order $n$ of fault-tolerance is increased. For $\Lambda$ to be well-defined, the system must operate with sufficiently low error rates to become more reliable as it grows. Note that it has been traditional to focus on a threshold error rate in the literature, as this is a theoretical quantity that is straightforward to calculate through simulation. The threshold error rate is not, however, a terribly meaningful experimental quantity, as its measurement would require variable error rate gates, necessitating the artificial insertion of noise. Rather, $\Lambda$ measures how far below threshold error rate a system is, without requiring an explicit measurement of this threshold value.

To obtain an estimate of $\Lambda$, one requires, at a minimum, a single system capable of demonstrating 1st- and 2nd-order fault-tolerance. It is not sufficient to compare a single qubit to the performance of 1st-order fault-tolerance as a bare qubit may have a very long memory time, but high-error gates, leading potentially to poorer performance of the error corrected system than a bare qubit, despite the system being below threshold ($\Lambda > 1$).

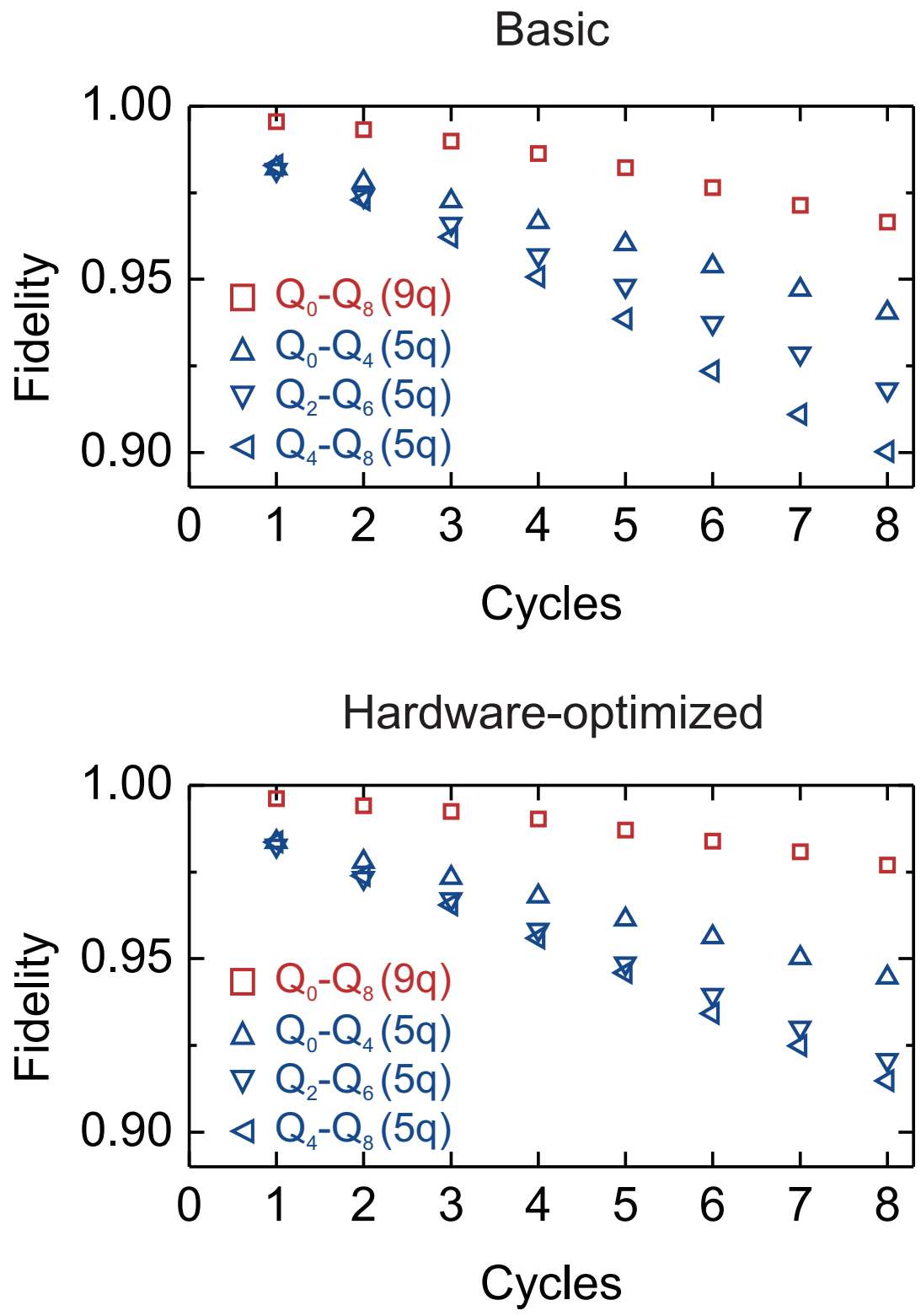

FIG. S14. **Comparison of basic data processing and hardware-optimized data processing.** (top) Fidelity of basic data processing for each combination of five qubits, and nine qubits $vs.$ cycle number. (bottom) Fidelity of hardware-optimized data processing for each combination of five qubits, and nine qubits $vs.$ cycle number.

In Fig. S14 we plot the three possible five qubit logical error rates and the nine qubit logical error rate for both basic and hardware-optimized postprocessing. To calculate $\Lambda$, we first calculate the average five qubit logical error rate and take the ratio of average five qubit error rate to nine qubit error rate. Given the accumulation of leakage, as discussed in the main text, we find an increase in error rate with cycle number, thus degrading $\Lambda$ with cycle number (Fig. S15). A truly scalable system would asymptote to a constant value of $\Lambda > 1$. Our data is currently insufficient to know if this will occur.

## V. PHYSICAL LEAKAGE MECHANISMS

From previous experiments, we measure a CZ gate to have $1-2\cdot10^{-3}$ population leakage, and a single qubit gates to have at most $\sim 10^{-4}$, determined in Ref.[5]. Empirically, we find that dispersive readout at higher powers can cause state transitions from $|1\rangle$ to $|2\rangle$, and we suspect this is the primary

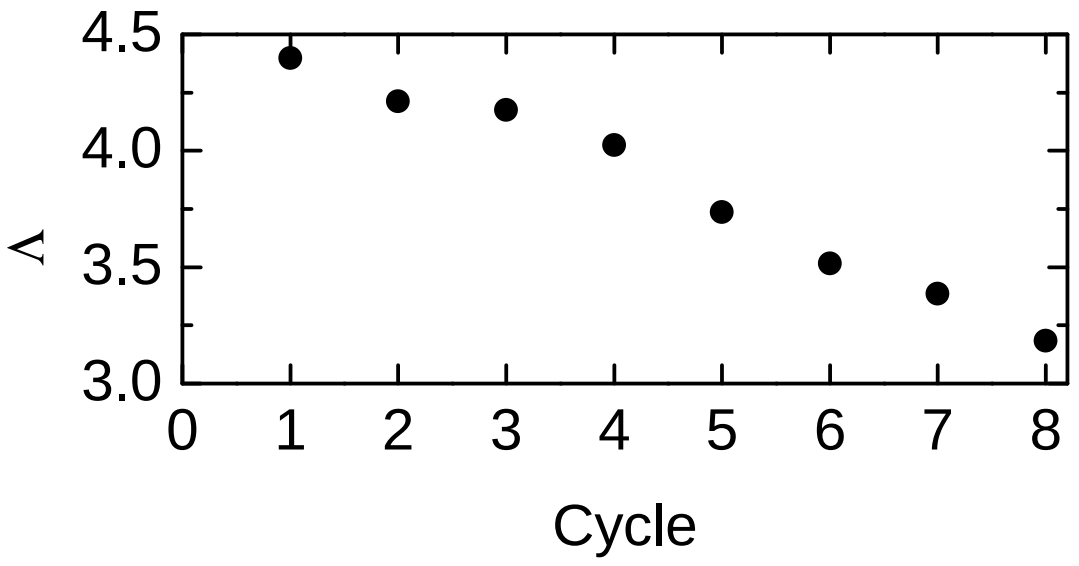

FIG. S15. **Changing $\Lambda$ with cycle number.** We find that $\Lambda$, the ratio of five qubit to nine qubit error rate, decreases with increasing cycle number, see section VII.

source of leakage.

## VI. DECOMPOSING THE FAILURE RATE

The state of the data qubits and the decomposition of the error rate is displayed in Fig. S16 for eight cycles of the nine qubit repetition code, with 90,000 statistics. The data qubit input state was $|00000\rangle$, hence any deviation of this state amounts to a bit error. In Fig. S16a, we find that the number of bit errors in the final data qubit state varies from a ratio of 0.33 for no error to a ratio of $6.9 \cdot 10^{-4}$ for five bit-flips. The ratio follows an exponentially decreasing trend with number of errors, indicating correlated bit-flips are nearly absent. We note that when running the repetition code for a large number of cycles, the physical state of the data qubits randomizes, and asymptotically approaches a uniform distribution.

At the end of each run, the minimum weight perfect matching algorithm gives one of two outcomes: the operator to flip the data qubits back to the input state (success), or to the bitwise inverse (failure). The error correction failure rate is plotted in Fig. S16b as a function of final data qubit state. We find no failure for the final data qubit state $|00000\rangle$. We find a failure rate ranging from $2 \cdot 10^{-3}$ to $11 \cdot 10^{-3}$ for the case of one bit-flip in the final data qubit state, and a general trend of increasing failure rate with increasing number of data qubit bit-flips.

The case for final state $|01011\rangle$ is highlighted, showing that in most cases the matching algorithm gives the correct operator $\hat{I} \otimes \hat{X} \otimes \hat{I} \otimes \hat{X} \otimes \hat{X}$, which changes the final state back to the input state. In 12.5% of these cases the matching algorithm gives the inverse operator, leading to failure. The final data qubit state may obscure measurement errors as well as multiple bit-flips which cancel each other.

## VII. INCREASING NUMBER OF DETECTION EVENTS WITH CYCLE

We observe an increasing pattern of detection events with cycle number. We attribute this to two main causes: state leakage, and measure qubit energy relaxation.

### A. State leakage

State leakage, the population of the non-computational $|2\rangle$-state, is shown to grow with repetition code cycle, see Fig. S17a. This population reduces the fidelity of the CZ entangling gate, leading to an increase in detection events.

In Fig. S17b, we plot the fraction of detection events, defined by the number of detection events divided by the total number of possible detection events, averaged over both input states and all eight-cycled runs. We plot this fraction when using standard $|0\rangle$ and $|1\rangle$ state discrimination (black), as well as when post-selecting out the detection events for $|2\rangle$ leakage (red).

For runs without $|2\rangle$-events, we notice I) a significantly reduced amount of detection events, II) nearly no dependence of detection events for measure qubits $Q_1$ and $Q_3$ on cycle number, III) a reduced dependence (slope) for qubit $Q_3$. The data suggest that non-computational leakage is a significant contribution to the amount of detection events and the increase with cycle number.

### B. Energy relaxation

We attribute the remainder of the increase in detection events to measure qubit energy relaxation, indirectly induced by randomization of data qubits. For input states $|0_L\rangle$ and $|1_L\rangle$, the measure qubits end in the state $|0\rangle$ at each cycle (see Fig. 1 of the main Letter). For our system $|0\rangle$ is highly robust. However, after several cycles some of the data qubits are likely flipped, and some of the measure qubits will switch between $|0\rangle$ and $|1\rangle$ with every round; $|1\rangle$ is susceptible to energy relaxation.

A clear indication of this error mechanism is shown in Fig. S17c, where we plot the fraction of detection events as a function of cycle number. The fraction of detection events alternate with every round. This is compatible with energy relaxation affecting the measure qubits, which switch between $|0\rangle$ and $|1\rangle$ with every round. The rise is then compatible with the increasing randomization of data qubits with cycle number. For either input state we find quantitatively a very similar behaviour, further supporting this interpretation.

## VIII. SAMPLE FABRICATION

Devices are fabricated identically to Ref.[5]. The fabrication details are reproduced here for convenience.

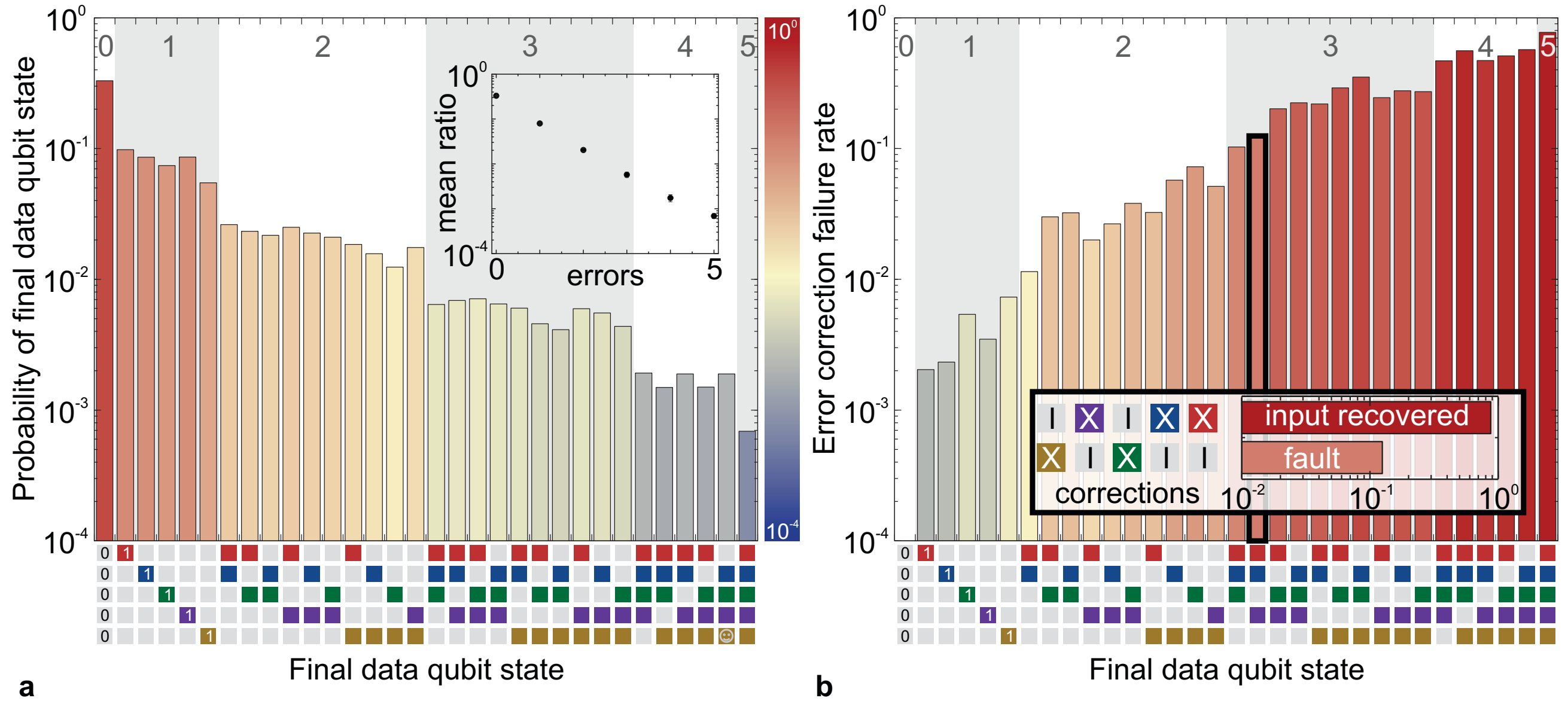

FIG. S16. **Decomposition of probability and failure rate for the nine qubit repetition code after eight cycles for input state $|00000\rangle$.** (a) Histogram of the final data qubit state, ordered with the number of bit-flips. Coloured squares indicate $|1\rangle$ states. Inset: averaged ratio versus number of bit-flips errors. (b) Error correction failure rate versus final data qubit state. The inset shows an example for the final data qubit state $|01011\rangle$ (highlighted): in 87.5% of the cases the minimum weight perfect matching gives the operator $\hat{I}\otimes\hat{X}\otimes\hat{I}\otimes\hat{X}\otimes\hat{X}$ which changes the final data qubit state into $|00000\rangle$, successfully recovering the input state. In 12.5% of the cases however, the inverse operator is given, leading to failure. We find no failure for the final data qubit state $|00000\rangle$. The experiment was repeated 90000 times.

The devices are made in a process similar to the fabrication steps outlined in Ref.[7], with an important improvement: we have added crossovers to suppress stray microwave chip modes by tying the ground planes together with low impedance connections. Otherwise, the many control wires in our chip could lead to segmentation of the ground plane, and the appearance of parasitic slotline modes[8]. The device is made in a five-step deposition process, (I) Al deposition and control wiring etch, (II) crossover dielectric deposition, (III) crossover Al deposition, (IV) Qubit capacitor and resonator etch, (V) Josephson junction deposition. The qubit capacitor, ground plane, readout resonators, and control wiring are made using molecular beam epitaxy (MBE)-grown Al on sapphire[9]. The control wiring is patterned using lithography and etching with a $BCl_3/Cl_2$ reactive ion etch. A 200 nm thick layer of $SiO_2$ for the crossover dielectric is deposited in an e-beam evaporator, followed by lift-off. We fabricate crossovers on all the control wiring, using a $SiO_2$ dielectric that has a non-negligible loss tangent. An in-situ Ar ion mill is used to remove the native $AlO_x$ insulator, after which a 200 nm Al layer for the crossover wiring is deposited in an e-beam evaporator, followed by lift-off. We used 0.9 $\mu$m i-line photoresist, lift-off is done in N-methyl-2-pyrrolidone at 80°C. A second $BCl_3/Cl_2$ etch is used to define the qubit capacitor and resonators; this step is separate from the wiring etch to prevent the sensitive capacitor from seeing extra processing. Lastly, we use electron beam lithography, an in-situ Ar ion mill, and double angle shadow evaporation to deposit the Josephson junctions, in a final lift-off process. See Ref.[7] for details.

## IX. QUBIT COHERENCE

We measure $T_1$ values for all nine qubits as a function of frequency in Fig. S18. We consistently find values in the 20-50 $\mu$s range.

In Fig. S19 we perform Ramsey experiments as a function of frequency. We find Ramsey $1/e$ times of 15 $\mu$s near the flux-insensitive point, and Ramsey $1/e$ times varying between 2 and 5 $\mu$s at frequencies away from the flux insensitive point.

## X. MEASUREMENT: READOUT AND BANDPASS FILTER DESIGN

The readout circuitry of the device is highlighted in Fig. S20. Qubits (blue) are coupled capacitively to individual readout resonators (purple). The readout resonators are coupled inductively to a bandpass filter (green), which is weakly coupled to the input (white, coupling quality factor $Q = 300$) and strongly coupled to the output (red, coupling quality factor $Q = 33$). At the output an impedance matched parametric amplifier (IMPA) acts as first stage amplifier[10]. The high bandwidth and saturation power is critical for system performance.

The bandpass filter isolates qubits from the 50 $\Omega$ environment formed by the readout lines. The bandpass filter (BPF) design used in this device is similar to Ref[6]. Here, the BPF is designed for a bandwidth of roughly 220 MHz, so placing nine readout resonators in this band with 30 MHz spacing requires a high level of precision. We use a design which ge-ometrically enforces minimal frequency difference between resonators and filter:

- The filter used here is a half wave ($\lambda/2$) resonator. This provides more space for coupling all qubits to the same filter.
- The input and output lines are coupled using voltage taps, which do not shift the filter frequency. A parallel plate capacitor, for example, could cause the frequency to shift if the dielectric thickness is unreliable.
- The coplanar waveguide geometry of the filter is chosen identical to that of the readout resonators. Therefore the kinetic inductance changes the frequencies of filter and resonators equally.
- The coupling capacitor between readout resonator and qubit has identical geometry for both data and measure qubits. This ensures a proper frequency spacing of readout resonators, as the self-capacitance, and therefore the pull on the readout resonator frequency, is identical.
- We chose a small resonator-filter coupling for data qubits to reduce measurement induced dephasing.
- We time-multiplex the readout of data and measure qubits to achieve high-fidelity readout for all nine qubits. The IMPA provides a gain of 15-18 dB, with saturation power around -100-110 dBm for the entire band. While the saturation power is high, it is a limitation for simultaneous readout of all nine qubits, necessitating time-multiplexing: We read out all five data qubits simultaneously, and all four measure qubits si-multaneously.

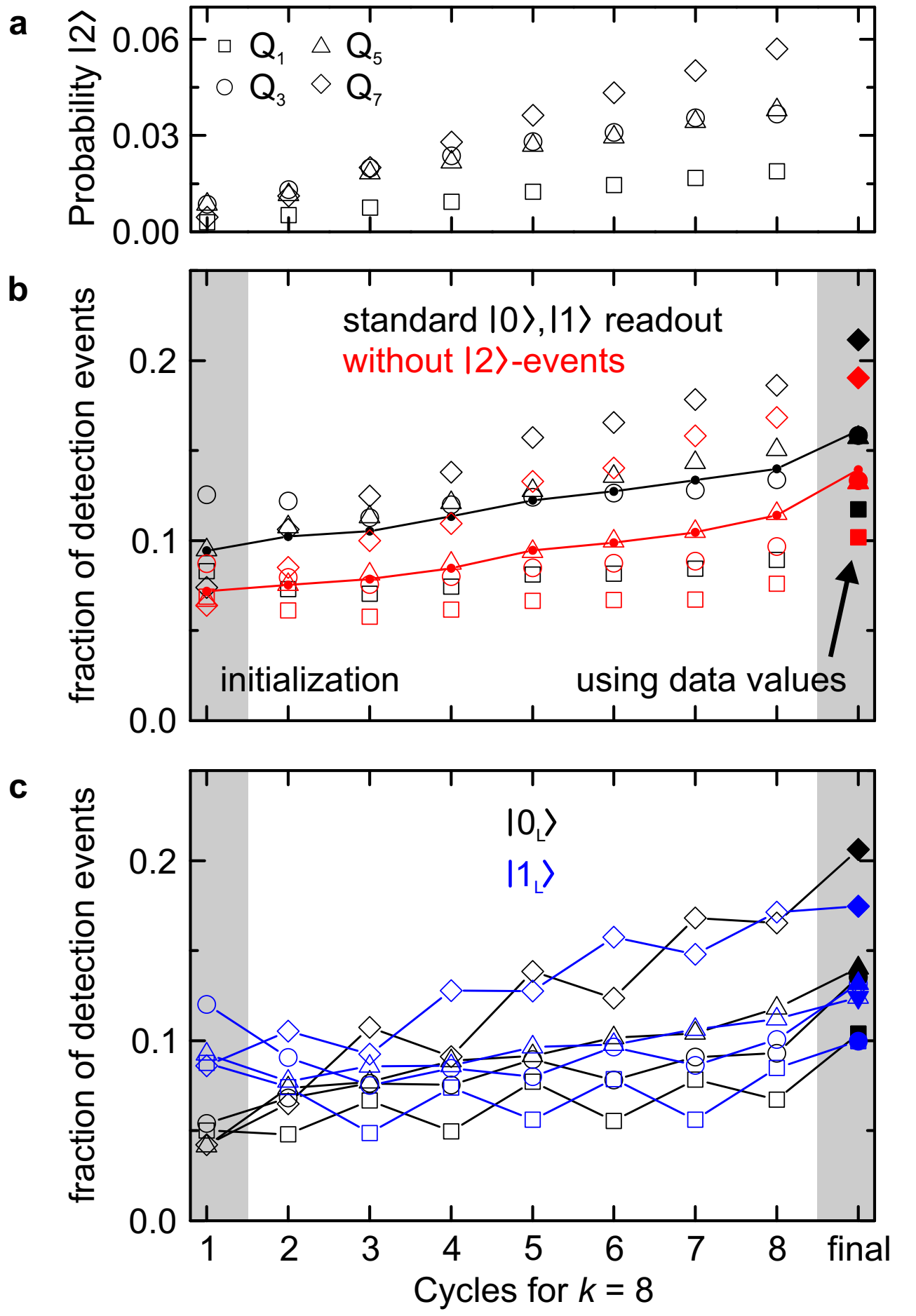

FIG. S17. **State leakage and fraction of detection events for $k = 8$.** (a) State leakage, $|2\rangle$-state population, for the four measurement qubits. Note that increase in measured $|2\rangle$ state population can come from misidentification of increasing $|1\rangle$ state population. (b) Fraction of detection events for the four measure qubits as a function of cycle number for standard $|0\rangle$ and $|1\rangle$ state discrimination (black), and without $|2\rangle$-events. Solid lines denote the averages. (c) Fraction of detection events for data qubit input states $|00000\rangle$ and $|11111\rangle$, showing a clear alternating pattern of increased detection events when measure qubits are likely in the $|1\rangle$ state. Used data is without $|2\rangle$-events.

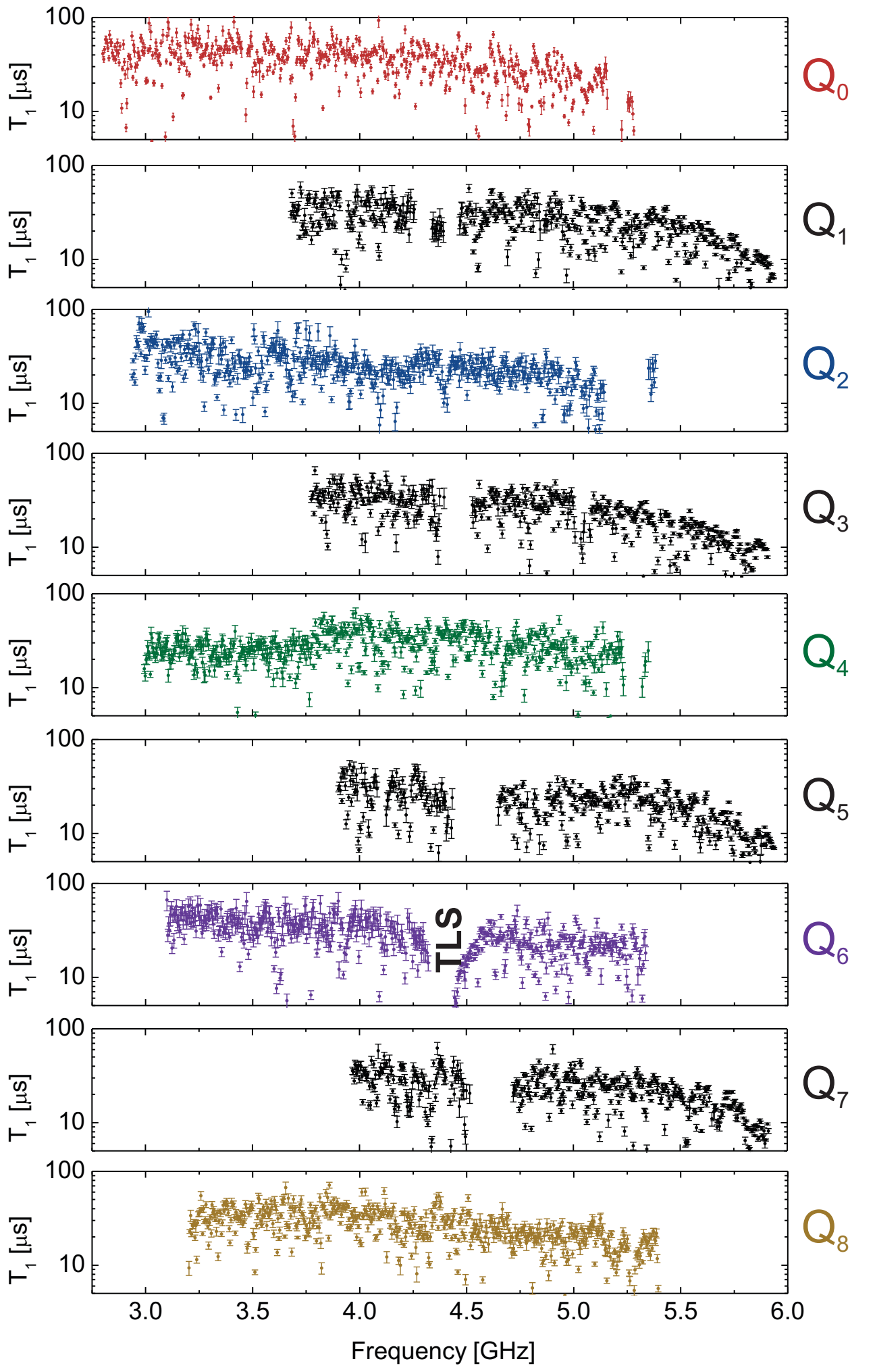

FIG. S18. **Energy Relaxation.** $T_1$ for all nine qubits as a function of frequency. Xmon qubits have a 24 $\mu$m gap and 24 $\mu$m center width for the capacitor. Missing data due to coherent swapping with other qubits or low quality fits. Two-level state (TLS) seen in the spectrum of $Q_6$.

## XI. MULTI-STATE READOUT

The demodulated inphase (I) and quadrature (Q) data from individual readout shots for preparing the $|0\rangle$, $|1\rangle$, and $|2\rangle$ states are shown in Fig S21 for each measure qubit. After accumulating statistics, the location in IQ space for the ideal $|0\rangle$, $|1\rangle$, and $|2\rangle$ states are determined. State discrimination is performed by calculating which ideal state is closest in IQ space to a measured value.

## XII. DEVICE PARAMETERS

The device parameters are listed in table S3. Note that the coupling rate $g$ is defined such that strength of the level splitting on resonance (swap rate) is $2g$ (Ref.[11]).

## XIII. PRESERVATION OF TWO-QUBIT GATE FIDELITY WHEN SCALING UP

We quantify the fidelity of multi-qubit operating by using two-qubit randomized benchmarking (RB), shown in Fig. S22. See Ref.[5] for details on the implementation of Clifford-based randomized benchmarking. The reference decay from performing two-qubit Cliffords $C_2$ is a metric for system performance, as it contains simultaneous single and two-qubit gates (each $C_2$ contains on average $\frac{33}{4}$ single qubit gates and $\frac{3}{2}$ CZ gates). We have performed two-qubit randomized benchmarking on qubits $Q_4$ and $Q_5$ as a character-istic pair. We find an average error per two qubit Clifford $C_2$ of 0.0191, which is close to the result of 0.0189 obtained for the five qubit chip in Ref.[5]. This shows that gate performance was maintained while scaling up to nine qubits and integrating high-fidelity measurement.

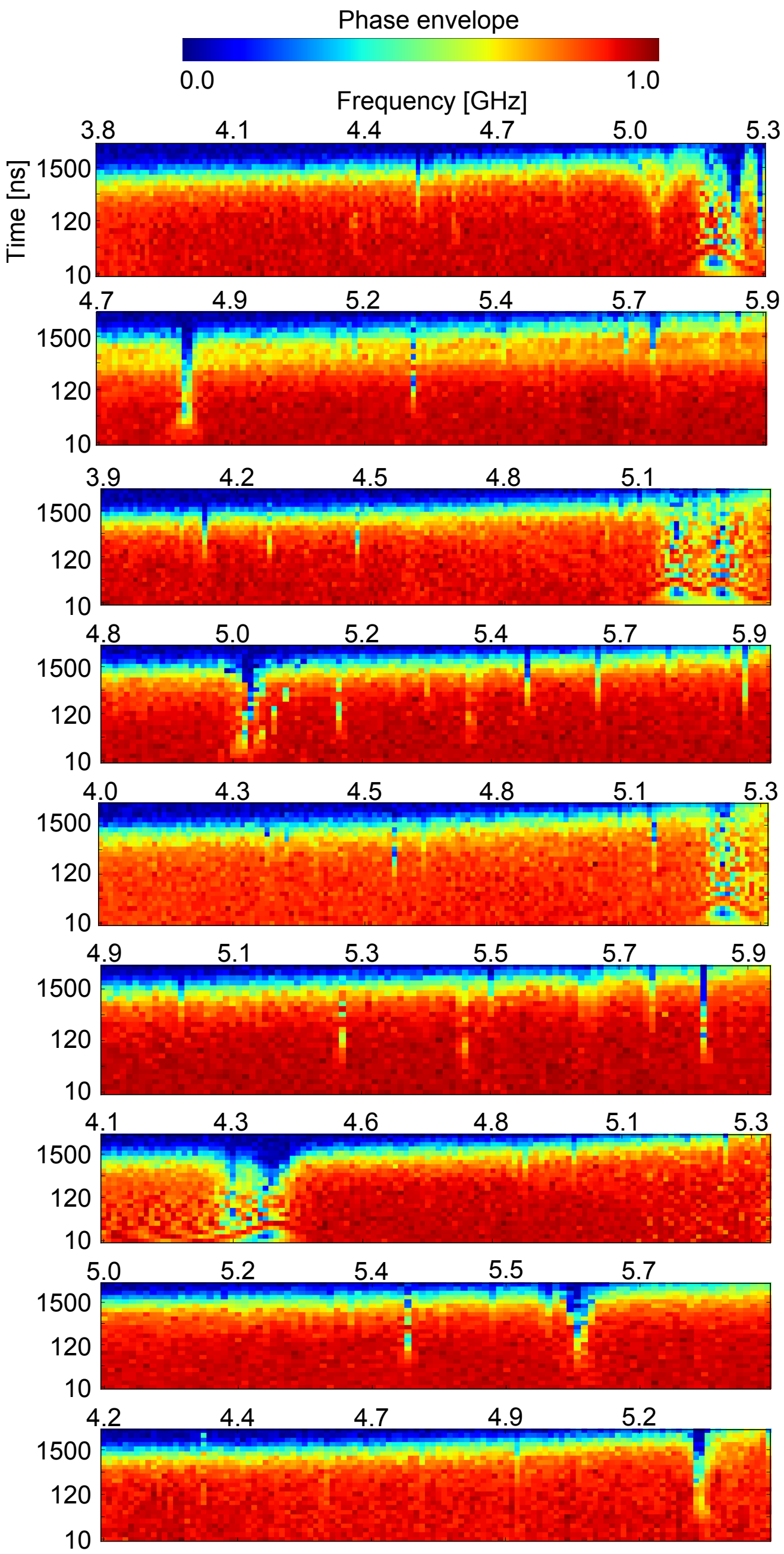

FIG. S19. **Dephasing.** Ramsey experiment for all nine qubits as a function of frequency. Phase envelope is the length of the Bloch vector on the XY plane. Oscillating features are interactions with coupled qubits.

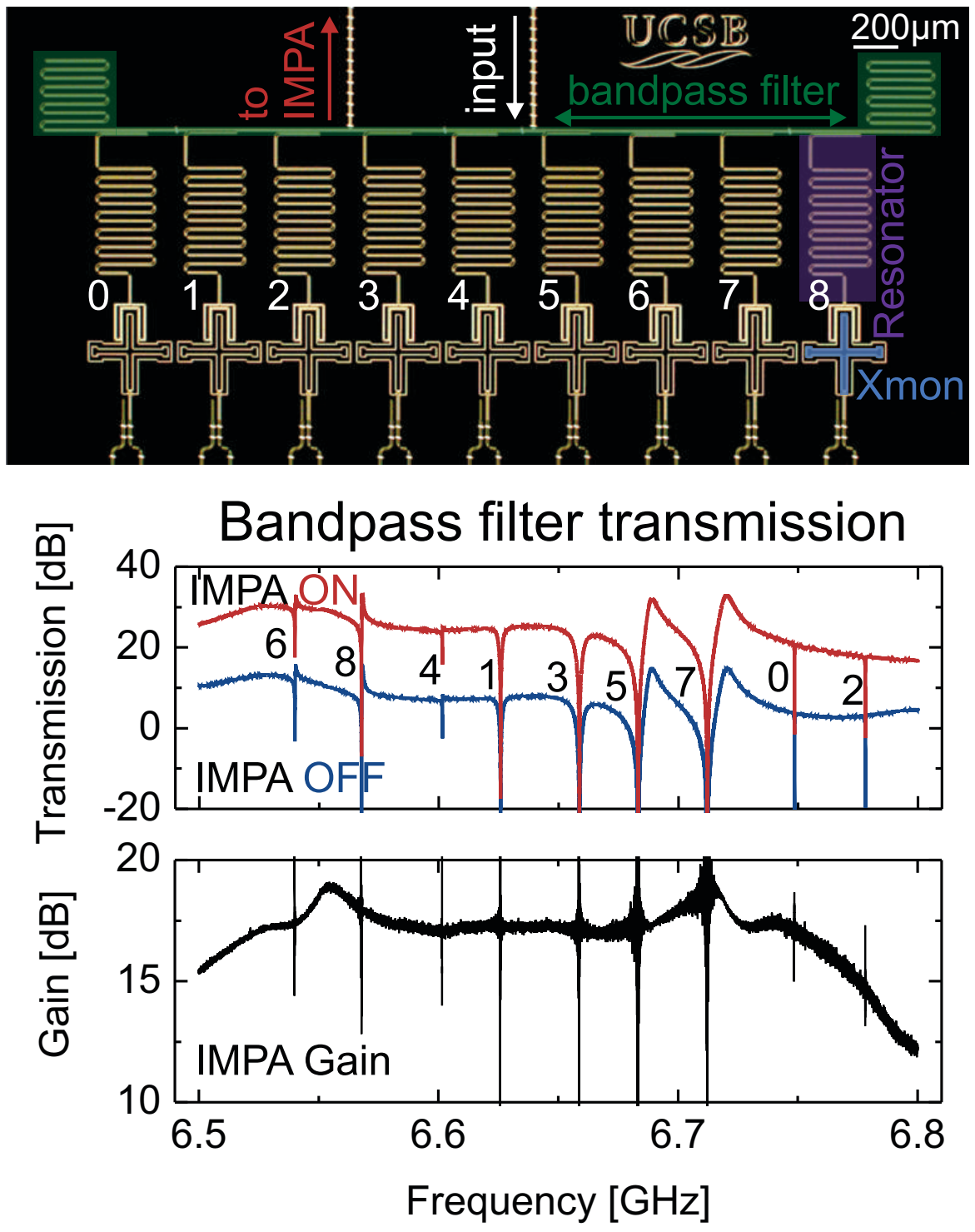

FIG. S20. **Bandpass filter and readout design.** Nine qubits (blue), with individual readout resonators (purple) are coupled to single half wave ($\lambda/2$) bandpass filter (green). The filter isolates qubits from the 50 $\Omega$ environment. A high bandwidth, high saturation power impedance matched parametric amplifier (IMPA) allows for frequency domain multiplexed readout using a single readout line.

TABLE S3. Parameters for the device when running the repetition code. $f$ are frequencies. $\eta$ is qubit nonlinearity. $g$ is coupling strength. $\kappa$ is resonator leakage rate.

| | $Q_0$ | $Q_1$ | $Q_2$ | $Q_3$ | $Q_4$ | $Q_5$ | $Q_6$ | $Q_7$ | $Q_8$ |
|---|---|---|---|---|---|---|---|---|---|
| **Qubit frequencies and coupling strengths** | | | | | | | | | |
| $f_{10}^{max}$ (GHz) | 5.30 | 5.93 | 5.39 | 5.90 | 5.36 | 5.94 | 5.33 | 5.91 | 5.39 |
| $\eta/2\pi$ (GHz) | -0.230 | -0.216 | -0.229 | -0.214 | -0.227 | -0.214 | -0.242 | -0.212 | -0.225 |
| $f_{10}^{idle}$ (GHz) | 4.3 | 5.18 | 4.43 | 5.28 | 4.49 | 5.40 | 4.60 | 5.46 | 4.7 |
| $f_{\mathrm{res}}$ (GHz) | 6.748 | 6.626 | 6.778 | 6.658 | 6.601 | 6.687 | 6.540 | 6.718 | 6.567 |
| $g_{\mathrm{res}}/2\pi$ (GHz) | 0.110 | 0.128 | 0.111 | 0.109 | 0.110 | 0.110 | 0.098 | 0.111 | 0.104 |
| $g_{\mathrm{qubit}}/2\pi$ (MHz) | 13.8 | | 14.1 | | 15.4 | | 14.4 | | |
| $g_{\mathrm{qubit}}/2\pi$ (MHz) | | 14.5 | | 14.4 | | 14.6 | | 15.6 | |
| $1/\kappa_{\mathrm{res}}$ (ns) | 675 | 69 | 555 | 30 | 1144 | 36 | 590 | 28 | 473 |
| **Readout (RO) parameters** | | | | | | | | | |
| RO error | 0.015 | 0.004 | 0.067 | 0.007 | 0.048 | 0.013 | 0.017 | 0.011 | 0.018 |
| simult. RO error | | 0.004 | | 0.012 | | 0.022 | | 0.013 | |
| separation error | | $4\cdot10^{-6}$ | | $2\cdot10^{-5}$ | | $2\cdot10^{-3}$ | | $2\cdot10^{-3}$ | |
| Thermal $\|1\rangle$ pop. | 0.013 | 0.007 | 0.028 | 0.01 | 0.037 | 0.018 | 0.012 | 0.009 | 0.012 |
| RO pulse length (ns) | 800 | 160 | 800 | 300 | 800 | 300 | 800 | 300 | 800 |
| RO demodulation length (ns) | 800 | 560 | 800 | 460 | 800 | 460 | 800 | 460 | 800 |
| $f_{10,\mathrm{RO}}$ (GHz) | | 5.46 | | 5.31 | | 5.40 | | 5.54 | |
| resonator $n_{\mathrm{photons}}$ | | 37 | | 18 | | 10 | | 14 | |
| **Gate parameters** | | | | | | | | | |
| Single qubit gate error | | 0.0006 | | 0.0009 | | 0.001 | | 0.001 | |
| $X_\pi$ length (ns) | 25 | 20 | 25 | 20 | 25 | 20 | 25 | 20 | 25 |
| CZ length (ns) | 45 | | 45 | | 45 | | 45 | | |
| CZ length (ns) | | 45 | | 45 | | 45 | | 45 | |
| **Qubit lifetime at idling point** | | | | | | | | | |
| $T_1$ ($\mu$s) | 26.3 | 24.7 | 39.2 | 21.3 | 41.1 | 19.1 | 22.0 | 28.1 | 18.6 |

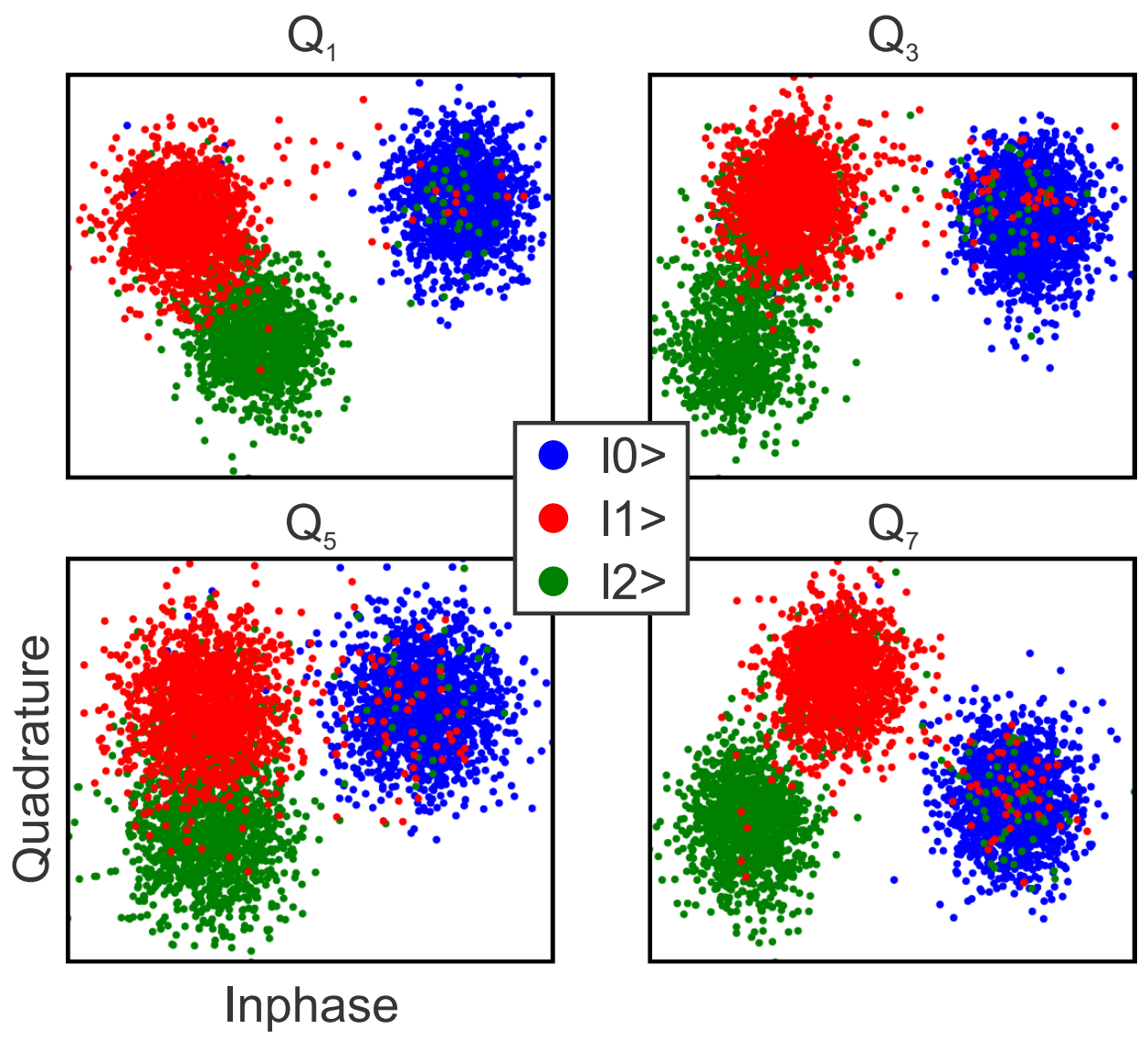

FIG. S21. **Three state readout for measure qubits.** Individual measurement shots for preparing the $|0\rangle$, $|1\rangle$, and $|2\rangle$ states for each measure qubit.

## XIV. MEASURE QUBITS IN DETAIL

We carefully characterize the four measure qubits in this device in Fig. S23. These qubits simultaneously combine long

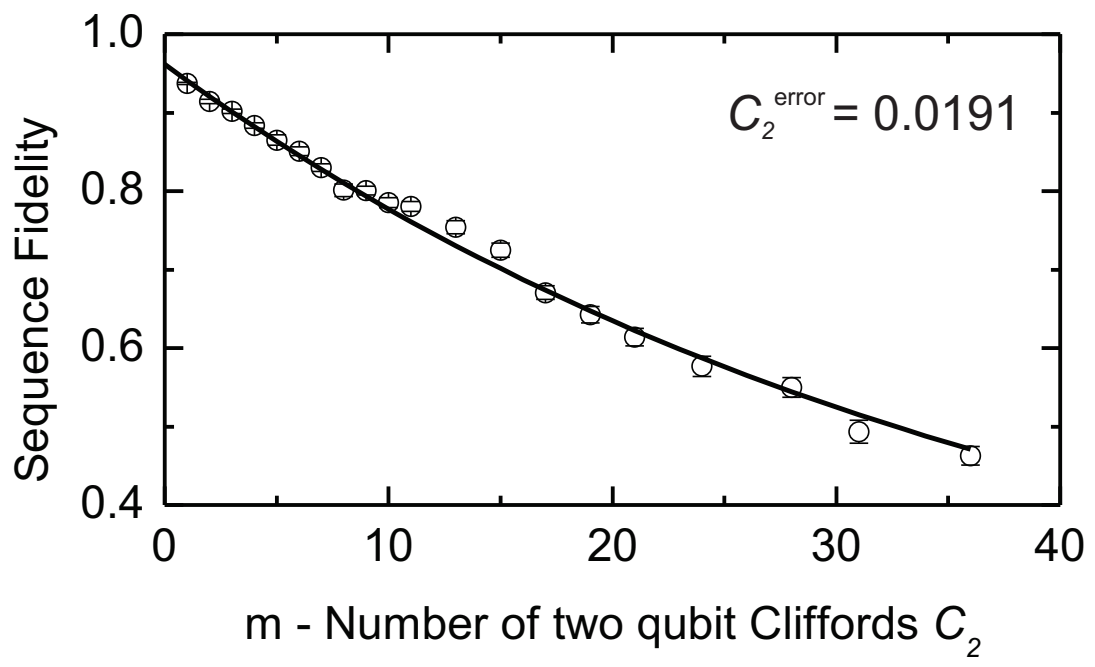

FIG. S22. **Two-qubit randomized benchmarking data.** The reference curve for two-qubit Cliffords $C_2$ for $Q_4$ and $Q_5$. The observed error rate is consistent with a simultaneous single qubit gate fidelity $> 0.999$ and a CZ gate fidelity $> 0.99$.

coherence with high fidelity gates and readout. For a full listing of parameters, see table S3.

## XV. Z CONTROL: CROSSTALK

We measure a crosstalk between the frequency Z control lines and qubits that is small, approximately $0.1 - 0.9\%$. The crosstalk matrix $M_\Phi$ is shown below, defined as: $\Phi_{\text{actual}} = (\mathbb{1} + dM_\Phi)\Phi_{\text{ideal}}$, with $\Phi$ the flux threaded through each qubit's superconducting quantum interference device (SQUID) loop. The magnitude of the crosstalk drops off with distance between lines and the sign depends on the routing direction of the wires, as expected.

$$dM_\Phi \approx 10^{-3} \begin{pmatrix} 0 & 2 & -1 & -2 & -3 & -1 & -2 & -2 & -2 \\ 2 & 0 & -5 & -2 & -3 & -1 & -2 & -1 & -2 \\ 2 & 5 & 0 & -5 & -3 & -2 & -3 & -1 & -2 \\ 1 & 4 & 2 & 0 & -7 & -3 & -3 & -2 & -2 \\ 1 & 4 & 2 & 2 & 0 & -8 & -5 & -2 & -2 \\ 1 & 3 & 2 & 2 & 0 & 0 & -9 & -4 & -2 \\ 1 & 3 & 1 & 2 & 0 & -1 & 0 & -7 & -2 \\ 1 & 2 & 1 & 2 & 0 & -1 & -5 & 0 & -6 \\ 1 & 2 & 1 & 2 & 1 & -1 & -7 & -1 & 0 \end{pmatrix}$$

## XVI. RUNNING THE REPETITION CODE

### A. Controlled-phase gates with nearest neighbor coupling

Our system consists of a linear array of qubits with nearest neighbor coupling $g$, where multiqubit $\Omega_{ZZ}$ interactions are turned on and off by frequency tuning[5]. As each qubit is coupled to more than one neighbor, operation frequencies must be carefully considered to prevent unwanted interactions.

There are three effects that must be simultaneously considered, see Fig. S24: I) The interaction which enables the CZ gate, by bringing the $|11\rangle$ and $|02\rangle$ close to resonance, needs to be turned off by detuning ($\Delta$) the qubits; this interaction turns of quadratically for $\Delta \gg g$. II) During a CZ gate, other neighboring qubits must be detuned to mitigate stray interactions, see Fig S24b. III) Next-nearest neighboring qubits have a small parasitic coupling (on the order of $g_{i,i+2}/2\pi = 0.75$ MHz). Therefore, while idling they effectively perform a parasitic CZ. To minimize this interaction we detune next-nearest neighbor qubits by $\eta/2 \approx 100$ MHz which is sufficient to prevent coupling of $|10\rangle$ to $|01\rangle$ and $|11\rangle$ to $|02\rangle$, see Fig. S24c.

With these three effects in mind, we consider how to perform the CZ portion of the repetition code. Each repetition code cycle consists of a CZ between each pair of neighboring qubits; 8 CZ gates for a 9 qubit array. This sequence can be executed, in principle, in the time of two CZ gates by performing 4 CZ gates in parallel at at time, this is technically demanding and resource intensive in terms of frequency space. Instead, we use a three step sequence which naturally mitigates stray interactions. Figure S24d shows the qubits in their idling state to minimize all interactions; neighboring qubits are detuned by $\Delta \approx 800$ MHz, next-nearest neighbor qubits are detuning by $\Delta \approx 100$ MHz. Additionally, this configuration is convenient to minimize microwave crosstalk between next-nearest neighbors, as resonant stray microwaves are detrimental to fidelity[13].

Figure S24e-g shows the three step operation to perform all eight CZ gates, where similar $\Delta$s are maintained between non-interacting qubits. Additionally, this basic pattern is scalable to a one-dimensional array of arbitrary length. We note that these essential techniques can be transferred to the operation of the two-dimensional surface code where each qubit has four nearest neighbors, instead of two.

### B. Evaluating the qubit spectrum

With generic operation principles outlined in the previous sections, we must choose specific operation frequencies for qubit idling, gates, and readout. In our Xmon qubits, the energy relaxation time $T_1$ varies over frequency; this is due to the fine structure from spectral distribution of incoherent defects[7]. Typically, $T_1$ times can differ by a factor of two to three over nearby frequencies, and careful characterization of the qubit spectrum is critical.

We see three kinds of features in the qubit spectrum. The most innocuous are incoherent defects, which suppress $T_1$ to the few $\mu$s range, but generally have small spectral width and can be safely swept past in frequency. We also see strongly and weakly coupled coherent features. Strongly coupled defects and strongly coupled qubits can coherently move population and affect phases – much more detrimental to algorithms – and sweeping qubits past them fast enough becomes challenging. Lastly, we see weakly coupled coherent defects and qubits which are slow enough such that we can move past them quickly in frequency. Thus, the qubits can be thought of as moving freely in a frequency band between strongly coherently coupled features, where operations should be done at frequencies away from incoherent defects and weakly coupled features.

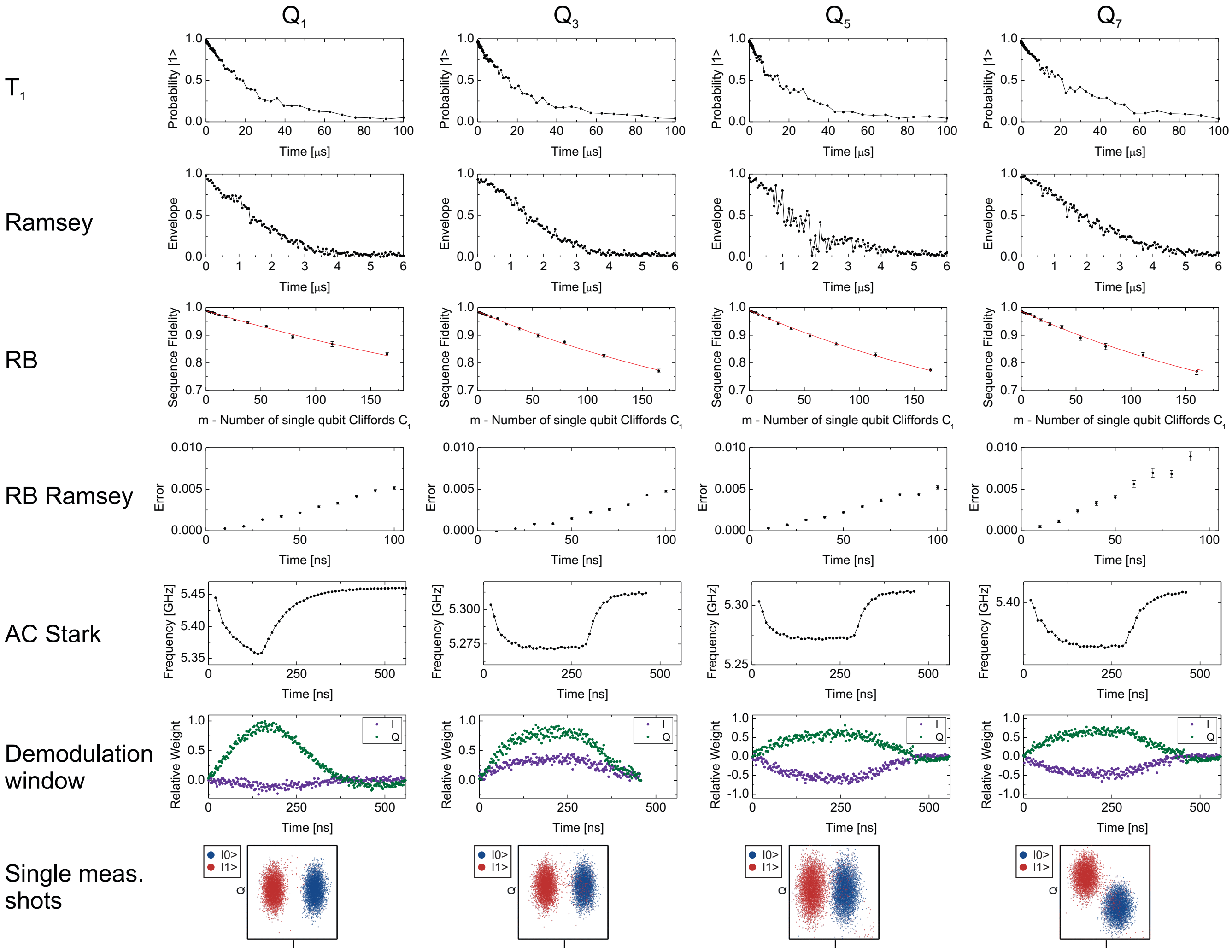

FIG. S23. **Measure qubit characterization.** We measure coherence, gate and readout properties on the four measure qubits. We find $T_1$ values in the 20-30 $\mu$s range for these operating points. Ramsey $1/e$ decay times are in the 2-3 $\mu$s range. Single qubit gate fidelities all meet or exceed 0.999. The RB Ramsey[12] experiment shows that the Ramsey 1/e time is not a limitation to gate fidelity. The AC stark shift and optimal demodulation windows displayed are used to achieve separation fidelities $> 0.998$.

Figure S25a shows the experiment for measuring the qubit spectrum. The qubit is excited to the $|1\rangle$ state, idles at a frequency for 100 ns, and measured. By sweeping over the operable frequency range for a qubit, we can identify population loss from incoherent or coherent features. Figure S25b shows the spectral data for the nine qubits. We identify bad regions in the spectrum by finding all frequencies that have population below a threshold value. We choose a threshold of 2% below the median $|1\rangle$ state population, corresponding to an added 5 $\mu$s $T_1$ mechanism over the qubit baseline. The data above threshold (operable regions) are plotted in black, and the data below threshold (inoperable regions) are plotted in red. Incoherent defects and weakly coupled coherent features (such as next-nearest neighbor qubits) have a thin spectral width and are easily avoided. Strongly coherently coupled features are easily identified through the coherent population swapping, such as seen in the spectrum of $Q_6$ at 4.3 GHz to 3.5 GHz and below.

### C. Programming the repetition code

With the principles for operating gates and efficiently using the qubit spectrum above, we now construct the repetition code algorithm. Figure S26a shows the high-level operation sequence for the nine qubit repetition code. Each measure qubit performs two CNOT gates with the control on the neighboring data qubits, and a measure. Figure S26a shows the actual physical gate sequence. For our system we decompose CNOT gates into CZ and $\pi/2$ gates. Additionally, we use detune gates to move unused qubits away in frequency space to avoid unwanted interactions, such as in Fig. S24. Echo $X_\pi$ gates are inserted between CZ and detune gates to suppress non-Markovian noise[12]. Figure S26b shows raw pulse waveforms for one repetition code cycle. The operations on each qubit have an XY, and Z control, as well as a multiplexed readout (RO) line. XY and RO waveforms are shown before up-conversion to GHz frequencies with an IQ mixer. Full control waveform data for eight cycles of the nine qubit repetition code can be found in Fig S27.

Operating frequencies for idling, readout, and CZ gates are chosen away from frequencies characterized to have poor coherence, as in section XVI B. Figure S28 shows qubit frequencies at various stages in the gate sequence. The vertical bar associated with each qubit shows operable (grey) and non-operable (red) frequencies. Note that during a CZ, the qubit higher in frequency has its $|2\rangle$ state virtually populated, thus it is important to this state to also avoid non-operable regions. We also plot the AC stark shift *vs.* time, as the qubit follows a non-trivial trajectory in frequency during readout, see Fig. S28.

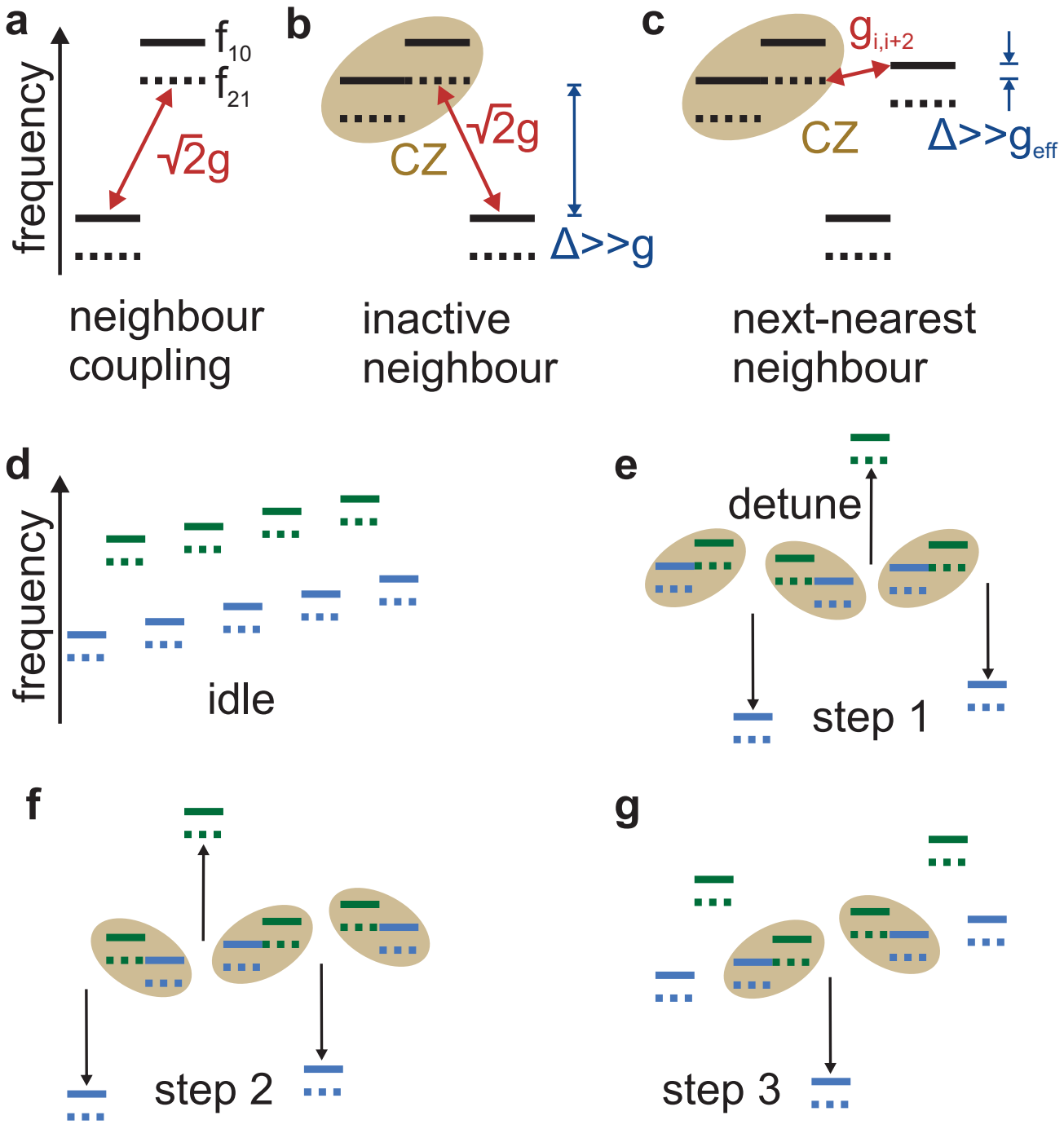

FIG. S24. **CZ interactions.** (a) Neighboring qubits are coupled with $\sqrt{2}g$ coupling between $|11\rangle$ and $|02\rangle$ states used for the CZ gate. (b) During a CZ gate, other neighboring qubits must be detuned to mitigate stray interactions. (c) Stray coupling between next-nearest neighbor qubits, approximately 1/20 of nearest neighbor coupling. A small detuning $\Delta$ is sufficient to avoid interaction. (d) Qubit idling frequencies to turn off nearest and next-nearest neighbor interactions. (e-g) Three step sequence to perform all 8 CZ of the repetition code with stray interactions in mind.

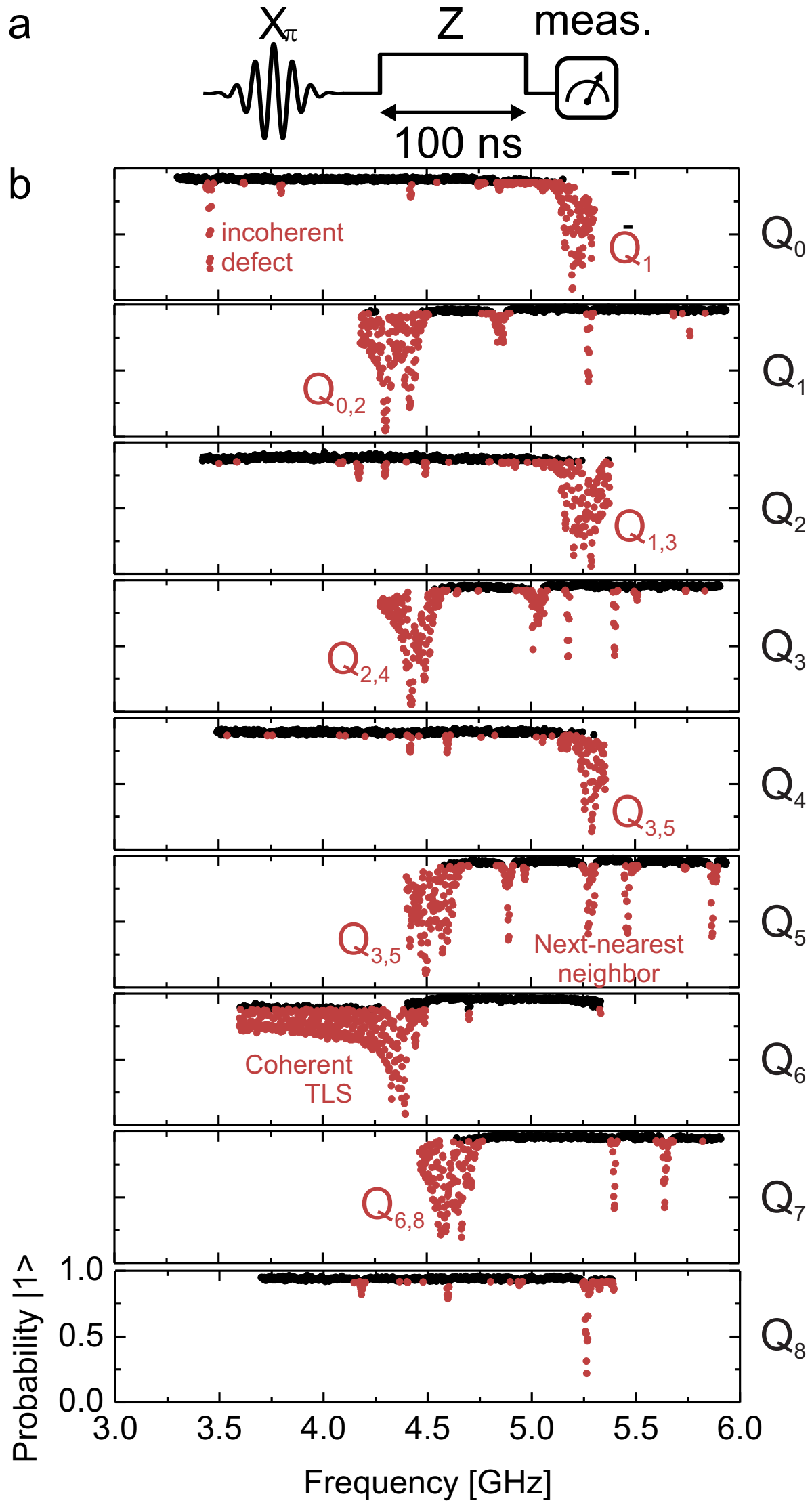

FIG. S25. **Qubit spectrum.** (a) Experiment for measuring the qubit spectrum. A qubit is excited to the $|1\rangle$ state and allowed to idle at different frequencies for 100 ns via a detuning pulse. (b) Spectrum data for nine qubits. Data is thresholded into good (black) and bad (red) regions. We define the threshold as 2% below the median $|1\rangle$ state population. This successfully identifies incoherent defects, coherent defects, and neighbouring as well as next-nearest neighbouring qubits.

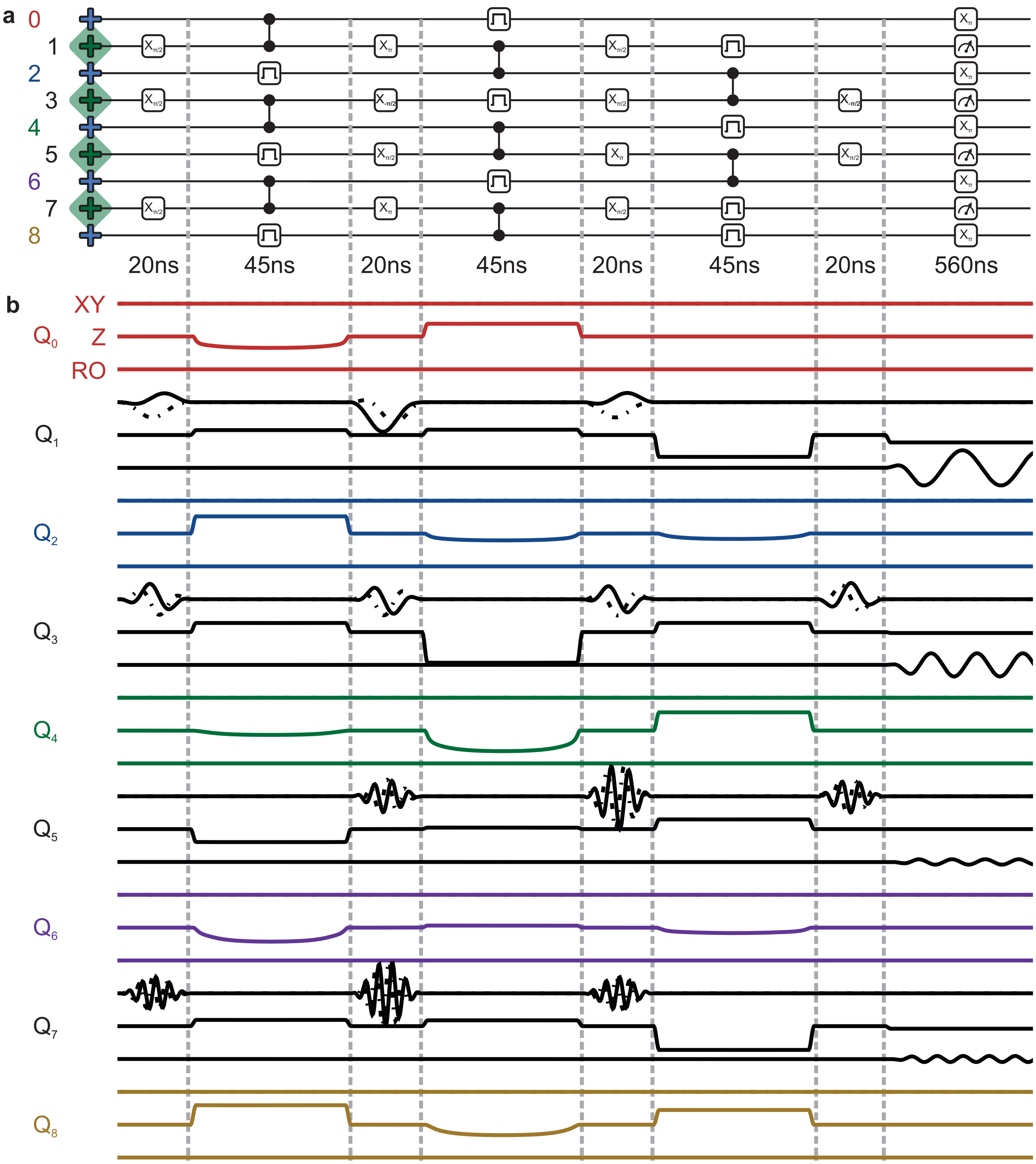

FIG. S26. **Single repetition code waveform data.** Rectangular gates indicate detuning pulses.

## XVII. EXPERIMENTAL SETUP

The experimental setup is described in Figure S29.

## XVIII. PROTECTING THE GHZ STATE FROM BIT-FLIP ERRORS: CONDITIONAL QUANTUM STATE TOMOGRAPHY

The density matrices of the GHZ states at the input and at the output of two cycles of the repetition code are reconstructed using quantum state tomography. We apply gates from $\{$ I, X/2, Y/2, X $\}^{\otimes 3}$. With quadratic maximum like-lihood estimation, using the MATLAB packages SeDuMi and YALMIP, we constrain the density matrix to be physical. Non-idealities in data qubit measurement and state preparation are suppressed by performing tomography on a zero-time idle[14,15]. For conditional tomography, we separate out the measured data qubit probabilities based on the detection events. The experiment was repeated $12 \cdot 10^3$ times.

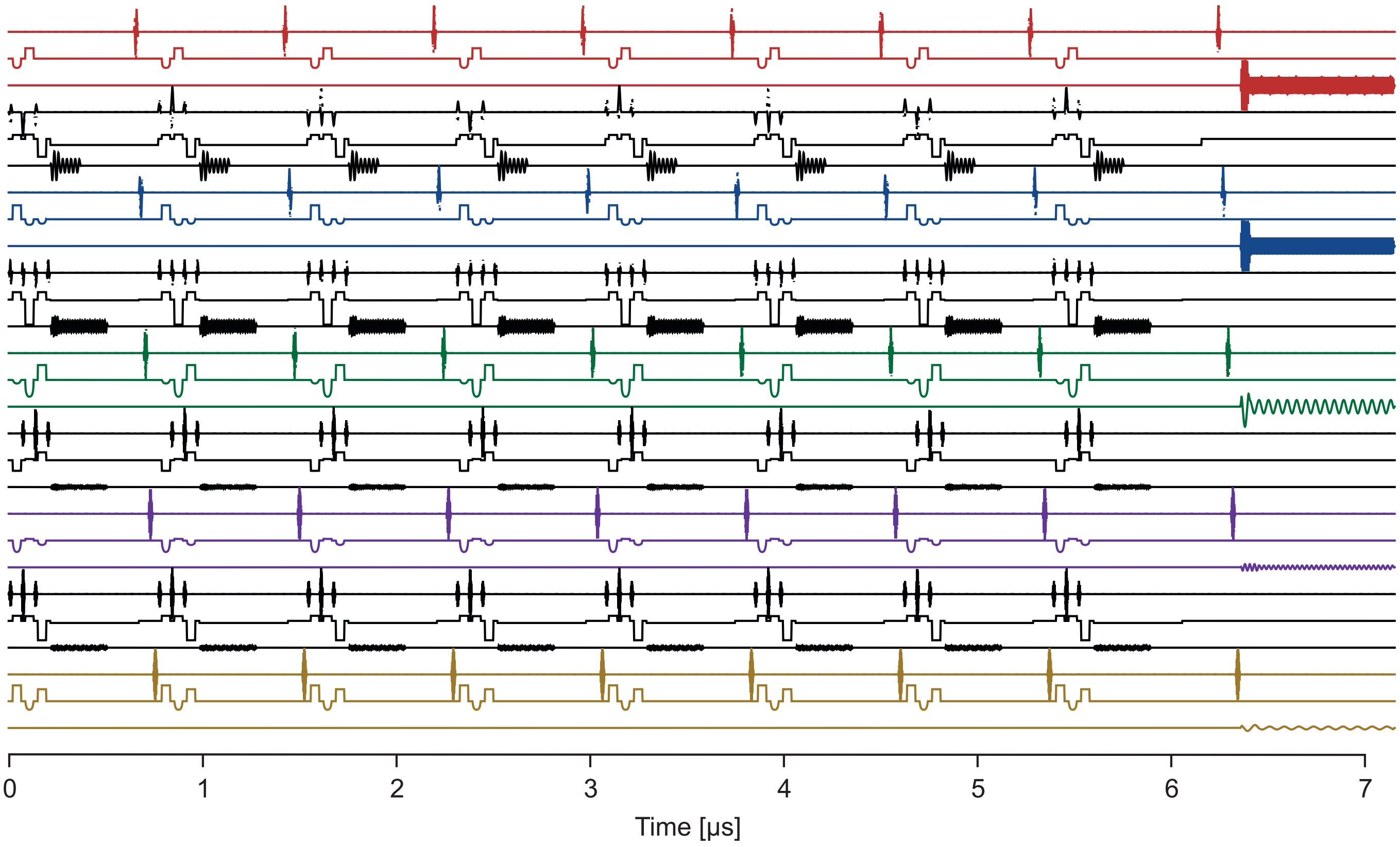

FIG. S27. **Waveform data for eight cycles of the nine qubit repetition code.**

The raw and corrected output density matrices are shown in Fig. S30.

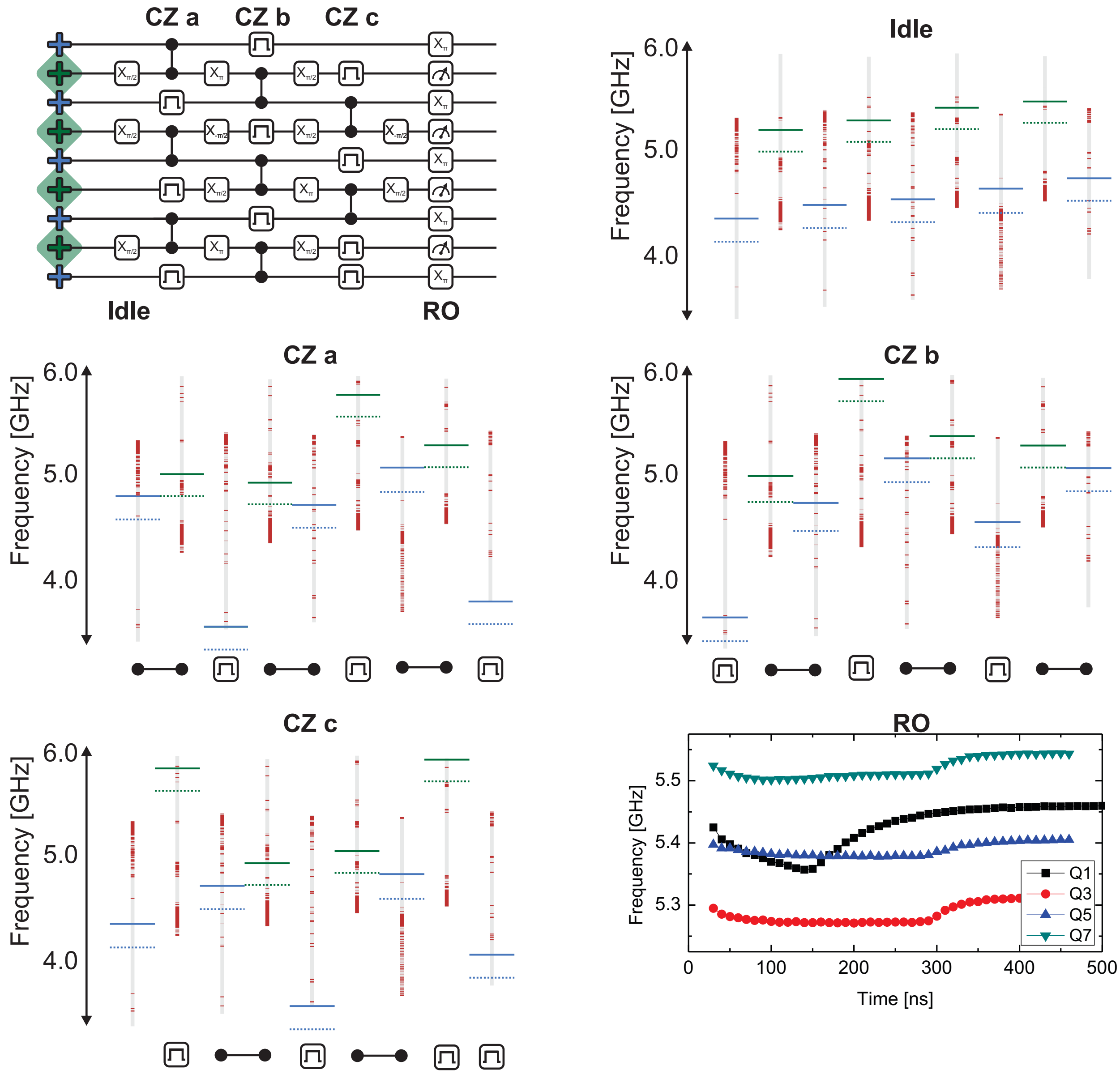

FIG. S28. **Qubit frequencies during the repetition code.** Vertical bars on qubit levels indicate operable regions (grey) and poor regions (red). Qubits are detuned to various frequencies during operation of the repetition code. Rectangular gates indicate detuning pulses. Different frequencies are chosen for idling, CZs and RO. Frequencies must be chosen to both reduce stray interactions (section XVI A) and microwave crosstalk, while also avoiding frequencies with poor coherence, see section XVI B.

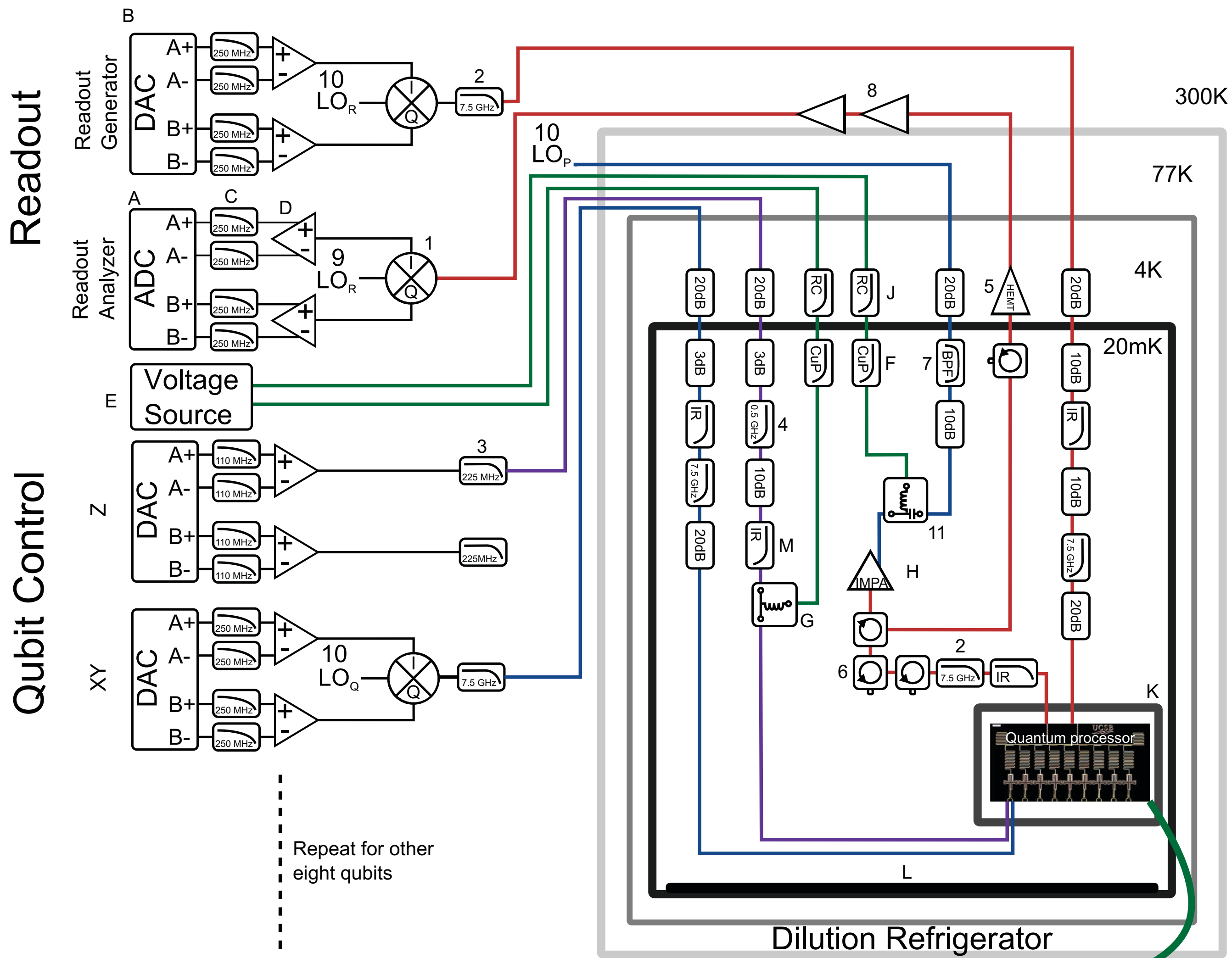

Commercial

| | |
|---|---|
| 1 | Marki IQ-0307 |
| 2 | Marki FLP-0750 |
| 3 | Mini Circuits VLFX-225 |
| 4 | Mini Circuits VLFX-500 |
| 5 | Low Noise Factory LNC4_8A |
| 6 | QuinStar CTH1392KS |
| 7 | Marki FB-1310 |
| 8 | Miteq AFS3-0010200-22-10P-4 |
| 9 | Hittite HMC-T2100 |
| 10 | Anritsu MG3692C |
| 11 | Sigatek SB11D2 |

Custom

| | |
|---|---|
| A | Analog to Digital Converter (ADC) |
| B | Digital to Analog Converter (DAC) |
| C | Gaussian Filter |
| D | Differential Amplifier |
| E | Voltage source ("Fastbias Card") |
| F | Copper powder & light tight LPF |
| G | DC Bias T |
| H | Parametric Amplifier (IMPA) |
| J | 1.5k cold resistor |
| K | Magnetic Shield |
| L | "IR-black" coating |
| M | Light tight LPF |

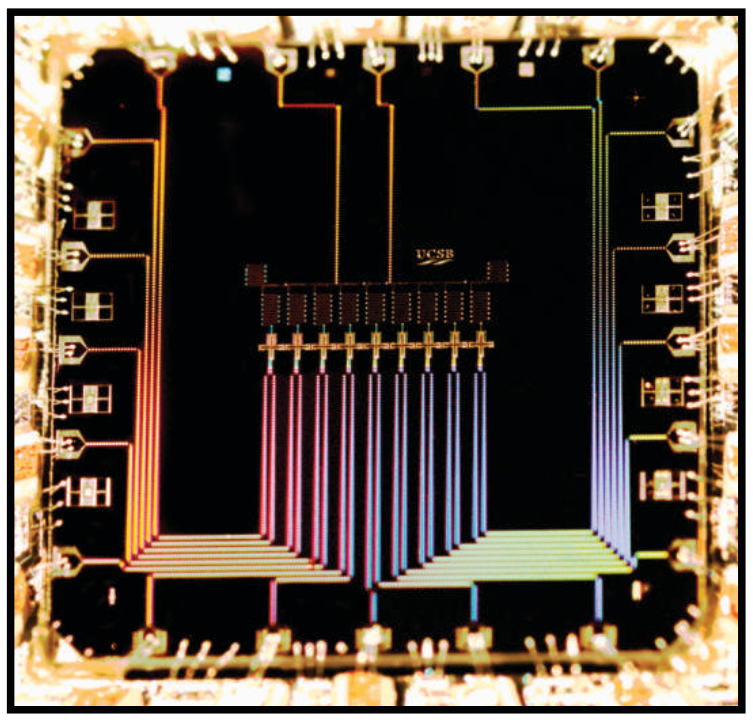

FIG. S29. **Electronics and Control Wiring.** Diagram detailing all of the control electronics, control wiring, and filtering for the experimental setup. Each qubit uses one digital to analog converter (DAC) channel for each of the X, Y, and Z rotations. Additionally, we use a DC bias tee to connect a voltage source to each qubit frequency control line to give a static frequency offset. All nine qubits are read out using frequency-domain multiplexing on a single measurement line. The readout DAC generates nine measurement tones at the distinct frequencies corresponding to each qubit's readout resonator. The signal is amplified by a wideband parametric amplifier[10], a high electron mobility transistor (HEMT), and room temperature amplifiers before demodulation and state discrimination by the analog to digital converter (ADC). All control wires go through various stages of attenuation and filtering to prevent unwanted signals from disturbing the quantum processor. Two local oscillators ($LO_Q$) are used for qubit XY control, at 4.38 and 5.202 GHz. The readout $LO_R$ is at 6.58 GHz. All LO, DAC, and ADC electronics are locked to a 10 MHz SRS FS725 rubidium frequency standard. Photograph of nine qubit device shown in the lower right.

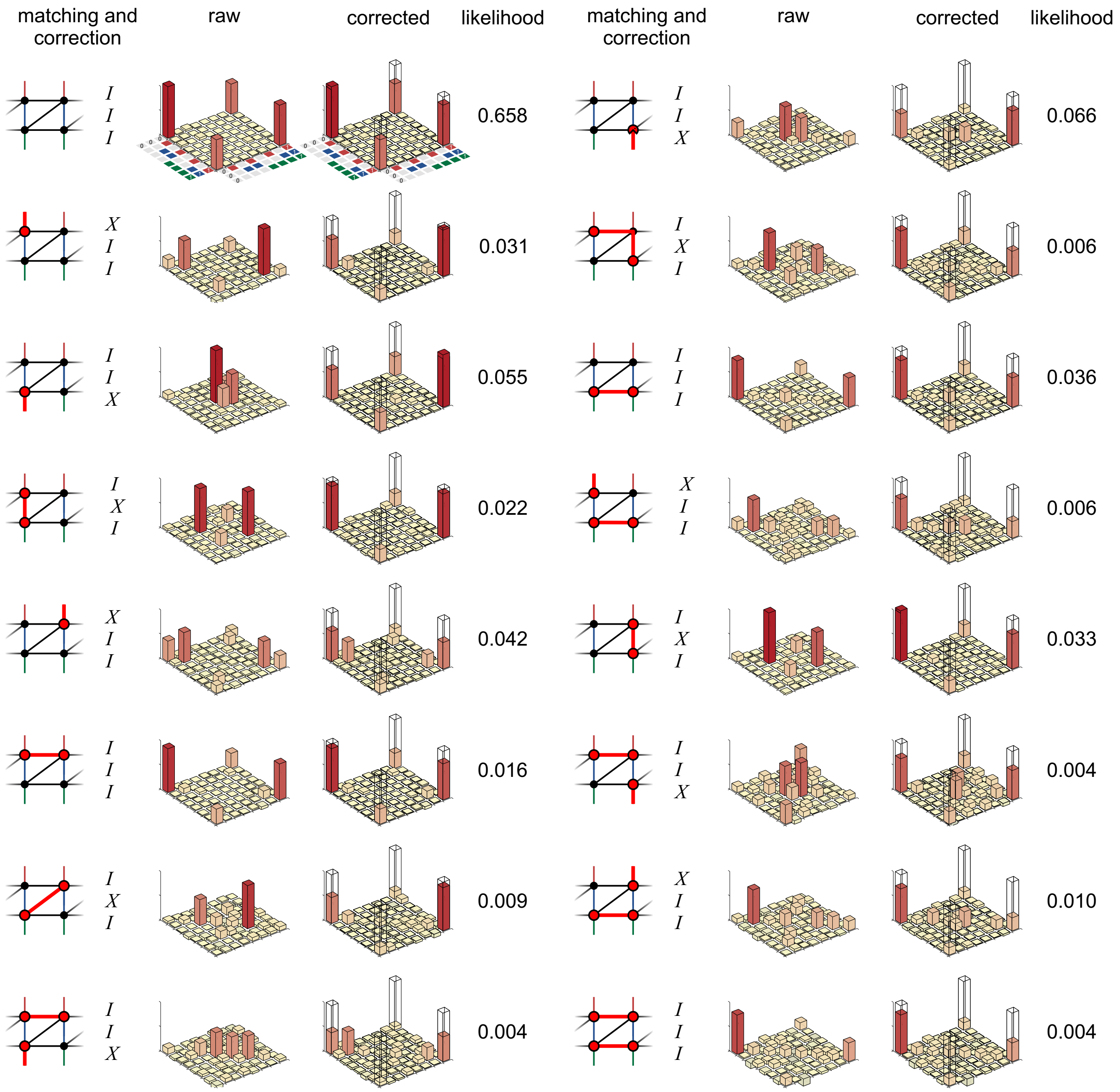

FIG. S30. **Protecting the GHZ state from bit-flip errors: All detection events.** The leftmost column displays the detection events and matching, indicating the most likely error, as well as the correction to apply in postprocessing to recover the input state. Raw and corrected output density matrices, reconstructed using quantum state tomography, conditional on all detection events. Corrected output density matrices are obtained by exchanging raw density matrix elements based on the correction. The real parts are shown. The likelihood indicates the prevalence of the detection event. See Fig. 3 of the main Letter for the quantum circuit diagram.

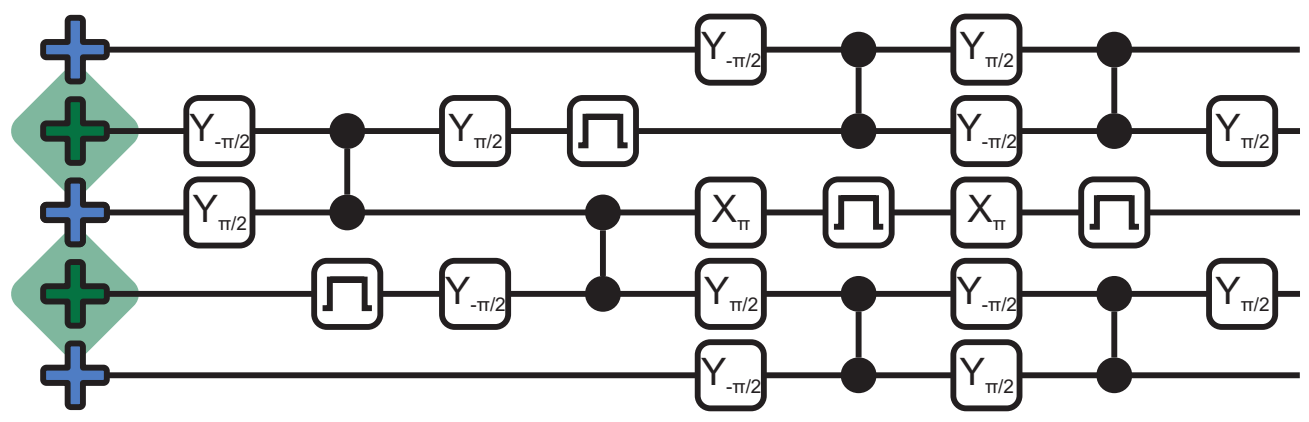

FIG. S31. **Quantum circuit for the generation of the GHZ state.** Rectangular gates indicate detuning pulses.

## XIX. GHZ GENERATION

The gate sequence for generation of the GHZ state used in the main text is shown in Fig. S31.